# Importance and methods to control, vary, and characterize mud strength for studying locomotion

Divya Ramesh, Gargi Sadalgekar, Qiyuan Fu, Zachary Souders, Jack Rao, *Chen Li

Department of Mechanical Engineering, Johns Hopkins University

*Corresponding author.

**Keywords:** terradynamics, flowable substrates, solid–fluid transition, solid–water mixture, granular media, penetrometer

**Abstract**

Animals and robots encounter mud at the water–land interface. Like sand, mud can stay solid or flow like a fluid. Unlike sand, the yield strength of mud at which solid–fluid transitions occur depends on not only the amount of solid relative to fluid (water in mud, air in dry sand), but also how much coarse grains and fine clay are within the solid. Despite understanding of locomotion on/within dry sand dominated by coarse grains with repulsive normal forces and friction, little is known for mud dominated by fine clay with strong cohesion. Here, we developed methods to prepare uniform mud of controlled, variable yield strength and characterize and track its drift from water evaporation. Compared to other flowable substrates, mud strength measured by upward force during penetration is weaker and can vary more, and mud sticks more during extraction to pull downward, making it more challenging for locomotion.

## 1. Introduction

Granular materials consisting of a collection of solid particles are ubiquitous in nature, and animals commonly and robots increasingly encounter them during locomotion. Locomotion on granular substrates is much more challenging than on solid ground, because they are "flowable" (Li et al., 2013). Although they normally stay solid, when the force exerted exceeds a critical threshold (called the yield strength), they flow like a fluid (Bonn et al., 2017; Jaeger and Nagel, 1992; Nedderman, 1992). Because of this solid–fluid transition, these substrates may suddenly yield and flow or re-solidify (Li et al., 2009; Mazouchova et al., 2010). Thus, an animal or robot's performance is sensitive to substrate yield strength (Li et al., 2009). Therefore, to systematically study and understand locomotion on flowable substrates, it is important to control, characterize, and vary substrate yield strength.

Granular substrates can have a wide range of solid composition and wetness (Coussot, 2017). They may consist of solid particles ranging from fine clay (up to ~10 μm) to coarse grains (~1 mm) or even pebbles and boulders (> ~1 cm) (Fig. 1A) and from completely dry to fully saturated (Coussot, 2017). Solid composition can be measured by fine clay fraction $\eta$ within the solid content (Coussot, 2017) (Fig. 1B, *x*-axis). Wetness can be measured by solid volume fraction $\phi$, the percentage of total volume occupied by the solid (Coussot, 2017) (Fig. 1B, *y*-axis). (Note that for dry sand, the volume not occupied by the solid is occupied by air, not water, and solid volume fraction $\phi$ measures the compaction of solid particles (Li et al., 2009), not wetness.)

Extensive research has advanced understanding of the bulk flow rheology of granular materials mixed with water as well as without, mostly for civil engineering and material handling applications (for a review, see (Coussot, 2017)). These materials can exhibit diverse mechanical behaviors depending on solid composition $\eta$ and wetness $\phi$ (Fig. 1B, a–f), because both affect yield strength at which solid–fluid transition occurs (Coussot, 2017). This classification of different regimes of dense granular flow rheology (Fig. 1B) provides a context for classifying and understanding research on animal and robot locomotion on granular substrates, which involve localized intrusion.

Based on their distinct rheology, we first grossly classify granular substrates (with or without water) into two broad types, depending on whether the solid content is dominated by coarse grains (sandy) or fine clay particles (muddy). (1) Sandy substrates with small $\eta$ (Fig. 1B, a–c), whose solid content is dominated by coarse grains (~1 mm). In this regime, repulsive, frictional contact forces between grains dominate solid–solid interactions. An intermediate amount of water between coarse grains can add weak cohesion via surface tension (Sharpe et al., 2015). (2) Muddy substrates with large $\eta$ (Fig. 1B, d–f), whose solid content is dominated by fine clay particles (up to ~10 μm). In this regime, colloidal effects from fine clay suspended in water can generate stronger cohesion (Coussot, 2017).

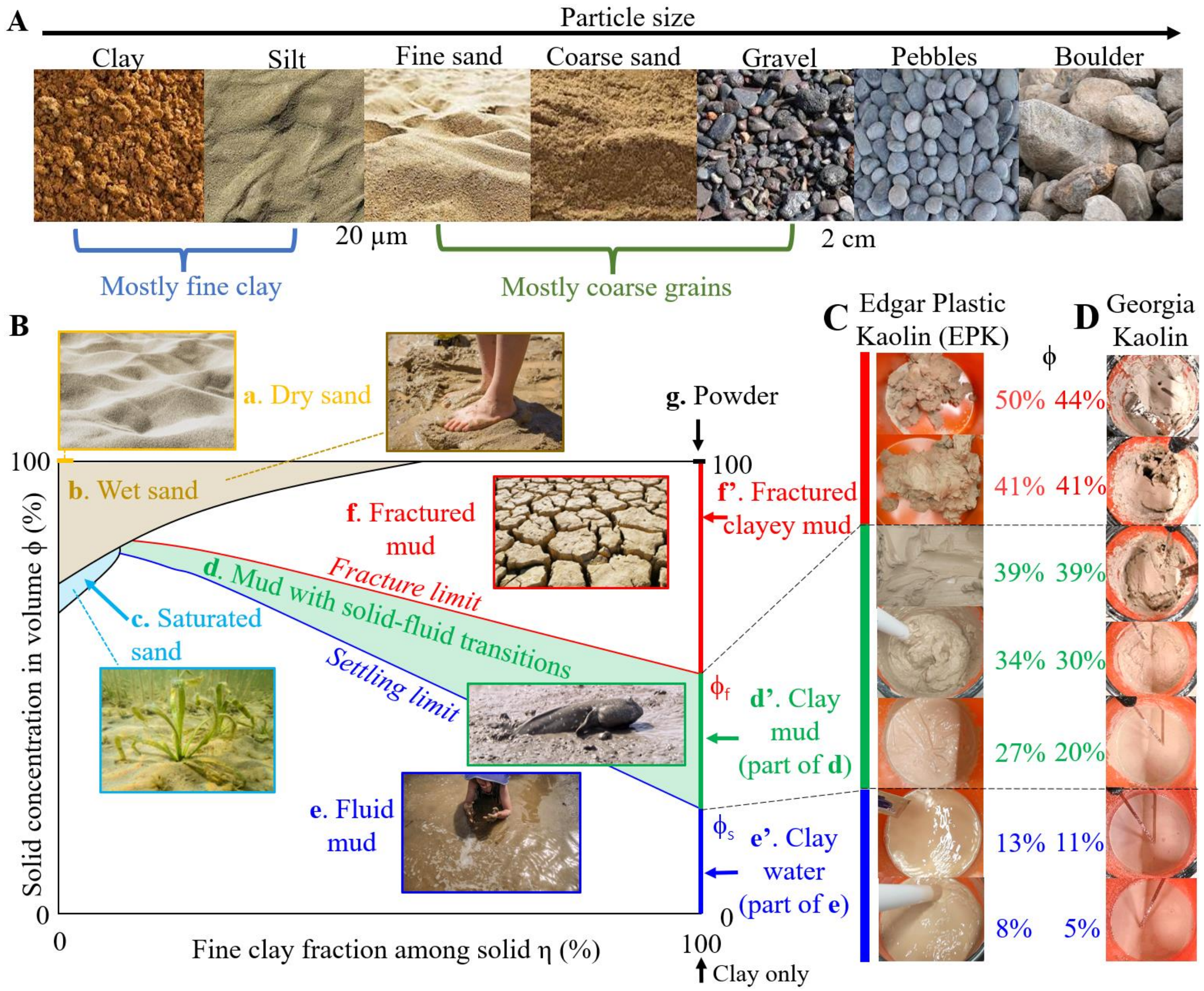

**Fig. 1. Classification of flowable substrates and controlled clay mud as a model wet flowable substrate. (A)** Granular particles of various sizes. **(B)** Diverse mechanical behavior of solid–water mixtures (and dry granular media, which are solid–air mixtures) fall in several regimes based on solid concentration measured by solid volume fraction $\phi$ and fine clay fraction $\eta$ among solid. Adapted from (Coussot, 2017). Note that in dry sand the volume not occupied by solid among the total volume is occupied by air, not water. **(C, D)** Clay mud (made from Edgar Plastic Kaolin (EPK) and Georgia Kaolin) with varied strength by varying solid volume fraction $\phi$. $\phi_f$ and $\phi_s$ are the upper and lower bounds of the solid–fluid transition

regime. Image Courtesy of (A) Sigur from shutterstock, tytyeu from iStock, Landscape Liquidator.CA, K.P Enterprises, Wikipedia, Meyer's turf and landscape nursery, and Southwest boulder & stone, (B, a–f) JOOINN, iStock (Elena Vafina), Dreamstime, Posterazzi, BBC, and Cavan Images.

We further divide each broad type into three sub-categories based on wetness $\phi$. Sandy substrates can be divided into: (a) Dry sand, with no water, i.e., $\eta = 0$ (Fig. 1B, a). (b) (Non-saturated) wet sand, with a little water insufficient to saturate (fill the space between) the grains, i.e., $\eta$ is small (Fig. 1B, b). (c) Water-saturated sand, with enough water to saturate the grains, i.e., $\eta$ is large (Fig. 1B, c). All these can go through solid–fluid transitions. Muddy substrates can also be divided: (d) Fractured solid mud with a large $\phi$ (Fig. 1B, f), which has so little water that it fractures and behaves like a solid. (e) Fluid mud with a small $\phi$ (Fig. 1B, e), which has so much water that most solid particles readily settle to the bottom, and the remaining, dilute, suspended fine clay makes the water viscous. (This mixture is thus a fluid and not a "substrate" per se.) (f) Mud with a medium $\phi$ capable of solid–fluid transitions (Fig. 1B, d), which has an intermediate level of water so that mud neither fractures nor settles. For a given $\eta$, the $\phi$ above which the mud factures is called the fracture limit ($\phi_H$, Fig. 1B), and the $\phi$ below which the mud settles is called the settling limit ($\phi_s$, Fig. 1B) (Coussot, 2017). (g) Clay (or other similarly small) particles that are completely dry can form a powder with strong cohesion due to Van der Waals forces or electrostatic interactions, which dominate gravity to form clumps and agglomerates (Kerimoglu et al., 2025). Among these seven sub-categories, only a, b, c, f are flowable substrates that can go through solid–fluid transition.

Whether granular particles can flow during locomotion also depends on particle size—which largely determines particle weight (particles in most natural flow substrates have a similar material density, ~2–3 × $10^3$ kg/$m^3$ (Coussot, 2017))—relative to the size of the animal or robot. This is because in general larger animals or robots can produce larger forces to displace the particles. Particles of comparable size to an animal or robot's appendage size or body width/height (e.g., pebbles, rocks, boulders) (Bergmann et al., 2017; Collins et al., 2013; Goodman, 2007; Jurestovsky et al., 2021; Mehta et al., 2021; Olberding et al., 2012; Parker and McBrayer, 2016; Redmann et al., 2020; Standen et al., 2016; Tomie et al., 2017; Tucker and McBrayer, 2012; Wehner, 2020) are likely too heavy to be easily pushed around. Thus, they present more of an uneven, mostly rigid (if slightly perturbable) terrain than flowable substrates.

Extensive research has advanced understanding of animal and robot locomotion on sandy substrates (Fig. 1B, regimes a–c; 82 studies in Table 1, 1–74, 80, 90–96), mostly on dry sand (Fig. 1B, regime a; 73 studies in Table 1, No. 3–17,19, 21–68, 70–74, 90, 92–93, 96), and some on wet (Fig. 1B, regime b; 13 studies in Table 1, No. 1–2, 8–9, 18, 26, 59–60, 80, 91–92, 94, 96) and saturated (Fig. 1B, regime c; 5 studies in Table 1, No. 20, 68–69, 90, 95) sand. Many of these studies controlled (51 studies in Table 1, No. 1, 3–16, 19, 21–23, 26, 28–30, 32, 35–40, 42–49, 51–55, 57–60, 72, 91–92, 96) and varied (17 studies in Table 1, No. 5, 8–9, 22–23, 26, 30, 35, 38, 44, 52–53, 55, 58–59, 92, 96) solid volume fraction $\phi$, and many studied forces on an intruder (representing animal or robot body and appendages) (57 studies in Table 1, No. 3–5, 7, 9, 11–17, 19, 22–23, 25, 27–28, 30–31, 33–36, 38–51, 54–61, 63, 65, 67–73, 92, 96). Such control, variation, and characterization of substrate yield strength revealed the neuromechanical principles of animal locomotion and enabled new capabilities and improved performance of robots in diverse sandy environments.

By contrast, an equivalent understanding of locomotion on and within muddy substrates is lacking, due to a lack of control, characterization, and variation for muddy substrates. There has been substantial research on locomotion on muddy substrates (Fig. 1B, regimes d–f; 17 studies in Table 1, No. 90–106), mostly in animals (12 studies in Table 1, No. 90–91, 93–95, 97–98, 100–101, 103, 105–106) and a few in robots (5 studies in Table 1, No. 92, 96, 99, 102, 104). A few studies (all in robots) systematically varied the wetness of clay mud (only fine clay, no coarse grains) (Table 1, No. 92, 96, 102, 104) and one varied the wetness of mud with both fine clay and coarse grains (Table 1, No. 79), and controlled $\phi$ of clay mud (Table 1, No. 92, 96, 99, 102, 104) and clay–grain mixtures (Table 1, No. 79). Only one animal walk and burrowing study varied $\phi$ for clay mud (Table 1, No. 106) and clay–grain mixtures (Table 1, No. 78), respectively, and three animal studies controlled clay mud wetness $\phi$ (Table 1, No. 91, 93, 101). However, no methods have been reported in detail in the literature showing exactly how to prepare large quantities of mud with controlled, variable properties and characterize its properties. Such control, variation, and characterization can be particularly useful because the wetness of mud can change from water evaporation, changing its properties.

Some animal studies used model “muddy” substrates that actually behave like a viscous fluid (Table 1, No. 87–89) or viscoelastic solid (Table 1, No. 80–86, 105), which likely do not go through solid–fluid transitions (to the best of our knowledge from reviewing these papers). Three fish locomotion studies used fine solid particles (Poly-Bore polymer, Table 1, No. 87; methyl cellulose, Table 1, No. 88, 89) and varied their solid concentration, but their $\phi$ (15% by volume) was most likely below or near the settling limit and not in the solid–fluid transition regime. These chemicals are well-known agents for creating viscous fluids (Horner and Jayne, 2014; Lutek and Standen, 2019; Lutek and Standen, 2021). This is further evidenced by the highest viscosity of Poly-Bore (Table 1, No. 87) and methyl cellulose (Table 1, No. 88–89) solutions used in these studies were 1000 mPa·s and 40 mPa·s respectively, which were 50% and 2% of the viscosity (2090 mPa·s) of clay mud (made from Georgia Kaolin clay) of $\phi$ = 27% near the lower bound of the solid–fluid regime (Fig. 1D). A few worm burrowing studies and a mudskipper study investigated the role of substrate mechanics (Table 1, No. 80–86, 105), but most of them used monolithic gelatin which is a viscoelastic solid that cannot go through solid–fluid transitions driven by mechanical forces (Bonn et al., 2017; Djabourov et al., 1988).

**Table 1: Literature review of flowable substrate control, variation, and characterization.** Varying and controlling ϕ are not applicable (N/A) for viscoelastic solids. Notes for Wetness level: “Fluid” means that it flows similar to mud below the settling limit. “Semi-fluid” means that it is in the solid–fluid transition regime but has more water content. “Semi-solid” means that it is in the solid–fluid transition regime but has more solid particles. “Solid” means that it behaves more like a rigid ground.

| No. | Reference | Substrate | Locomotor | Locomotor mode/movement type | Wetness level | Varied ϕ | Controlled ϕ | Studied locomotor forces | Studied kinematics | Studied behavior | Studied motor control | Studied gait transition |
|---|---|---|---|---|---|---|---|---|---|---|---|---|
| 1 | (Mehta et al., 2021) | Sand, Pebbles | Animal | Lateral bending | Wet | No | Yes | No | Yes | Yes | Yes | Yes |
| 2 | (Redmann et al., 2020) | Sand, Pebbles | Animal | Lateral undulation | Wet | No | No | No | Yes | Yes | Yes | No |
| 3 | (Agarwal et al., 2019) | Sand | Probe | Wheeled | Dry | No | Yes | Yes | No | Yes | No | No |
| 4 | (Agarwal et al., 2021a) | Sand | Probe | Wheeled | Dry | No | Yes | Yes | Yes | Yes | No | No |
| 5 | (Aguilar and Goldman, 2016) | Sand | Robot | Jump | Dry | Yes | Yes | Yes | Yes | Yes | Yes | No |
| 6 | (Austin et al., 2022) | Sand | Robot | Jump | Dry | No | Yes | No | Yes | Yes | Yes | No |
| 7 | (Ozkan Aydin et al., 2017) | Sand | Animal, Robot | Walk | Dry | No | Yes | Yes | Yes | Yes | No | No |
| 8 | (Bagheri et al., 2017) | Sand | Animal, Robot | Run | Dry, wet | Yes | Yes | No | Yes | Yes | No | No |
| 9 | (Bagheri et al., 2023) | Sand | Robot | Walk, run | Dry, wet | Yes | Yes | Yes | Yes | Yes | Yes | No |
| 10 | (Bagheri et al., 2024) | Sand | Robot | Burrow | Dry | No | Yes | No | Yes | Yes | No | No |
| 11 | (Chang et al., 2021) | Sand | Robot | Jump | Dry | No | Yes | Yes | Yes | Yes | Yes | No |
| 12 | (Chong et al., 2018) | Sand | Animal, Robot | Walk | Dry | No | Yes | Yes | Yes | Yes | Yes | No |
| 13 | (Chong et al., 2021) | Sand | Animal, Robot | Walk | Dry | No | Yes | Yes | Yes | Yes | Yes | No |
| 14 | (Chopra et al., 2020) | Sand | Probe | Drop | Dry | No | Yes | Yes | Yes | Yes | No | No |
| 15 | (Chopra et al., 2023) | Sand | Robot | Swim | Dry | No | Yes | Yes | Yes | Yes | No | No |
| 16 | (Ding et al., 2011) | Sand | Probe | Drag | Dry | No | Yes | Yes | No | Yes | No | No |
| 17 | (Ding et al., 2013) | Sand | Animal, Robot | Swim | Dry | No | No | Yes | Yes | Yes | Yes | No |
| 18 | (Dorgan, 2018) | Sand | Animal | Burrow | Wet | No | No | No | Yes | Yes | No | No |
| 19 | (Gart et al., 2021) | Sand | Robot | Run | Dry | No | Yes | Yes | Yes | Yes | Yes | No |

| No. | Reference | Substrate | Locomotor | Locomotor mode/movement type | Wetness level | Varied ϕ | Controlled ϕ | Studied locomotor forces | Studied kinematics | Studied behavior | Studied motor control | Studied gait transition |
|---|---|---|---|---|---|---|---|---|---|---|---|---|
| 20 | (Gidmark et al., 2011) | Sand | Animal | Burrow | Saturated | No | No | No | Yes | Yes | No | No |
| 21 | (Gosyne et al., 2018) | Sand | Robot | Walk | Dry | No | Yes | No | Yes | Yes | Yes | No |
| 22 | (Gravish et al., 2010) | Sand | Probe | Drag | Dry | Yes | Yes | Yes | Yes | Yes | No | No |
| 23 | (Gravish et al., 2014) | Sand | Probe | Drag | Dry | Yes | Yes | Yes | No | Yes | No | No |
| 24 | (Guo et al., 2021) | Sand | Robot | Bounding | Dry | No | No | No | Yes | Yes | No | No |
| 25 | (Hall et al., 2022) | Sand | Animal | Hop | Dry | No | No | Yes | Yes | Yes | Yes | No |
| 26 | (Han et al., 2023) | Sand | Animal | Walk | Dry, wet | Yes | Yes | No | Yes | Yes | Yes | No |
| 27 | (Hatton et al., 2013) | Sand | Robot | Swim | Dry | No | No | Yes | Yes | Yes | Yes | No |
| 28 | (He et al., 2023) | Sand | Probe | Anchor intrusion | Dry | No | Yes | Yes | No | Yes | No | No |
| 29 | (Huang and Tao, 2022) | Sand | Robot | Burrow | Dry | No | Yes | No | Yes | Yes | No | No |
| 30 | (Huang et al., 2020) | Sand | Robot | Burrow | Dry | Yes | Yes | Yes | Yes | Yes | No | No |
| 31 | (Huang et al., 2022) | Sand | Robot | Burrowing | Dry | No | No | Yes | Yes | Yes | No | No |
| 32 | (Hubicki et al., 2016) | Sand | Robot | Jump | Dry | No | Yes | No | Yes | Yes | Yes | No |
| 33 | (Huh et al., 2023) | Sand | Robot | Walk-burrow-tug | Dry | No | No | Yes | Yes | Yes | No | Yes |
| 34 | (Jafarnezhadgero et al., 2022) | Sand | Animal | Run | Dry | No | No | Yes | Yes | Yes | No | No |
| 35 | (Li et al., 2009) | Sand | Robot | Walk | Dry | Yes | Yes | Yes | Yes | Yes | No | Yes |
| 36 | (Li et al., 2010a) | Sand | Robot | Walk | Dry | No | Yes | Yes | Yes | Yes | Yes | No |
| 37 | (Li et al., 2010b) | Sand | Robot | Walk | Dry | No | Yes | No | Yes | Yes | No | No |
| 38 | (Li et al., 2012a) | Sand | Probe | Drag | Dry | Yes | Yes | Yes | Yes | Yes | No | No |
| 39 | (Li et al., 2012b) | Sand | Animal | Run | Dry | No | Yes | Yes | Yes | Yes | Yes | No |
| 40 | (Li et al., 2013) | Sand | Robot | Walk | Dry | No | Yes | Yes | Yes | Yes | No | No |
| 41 | (Li et al., 2021) | Sand | Robot | Swim | Dry | No | No | Yes | Yes | Yes | No | No |
| 42 | (Lynch et al., 2020) | Sand | Robot | Hop | Dry | No | Yes | Yes | Yes | Yes | Yes | No |

| No. | Reference | Substrate | Locomotor | Locomotor mode/movement type | Wetness level | Varied ϕ | Controlled ϕ | Studied locomotor forces | Studied kinematics | Studied behavior | Studied motor control | Studied gait transition |
|---|---|---|---|---|---|---|---|---|---|---|---|---|
| 43 | (Lynch et al., 2025) | Sand | Robot | Hop | Dry | No | Yes | Yes | Yes | Yes | No | No |
| 44 | (Maladen et al., 2009) | Sand | Animal | Swim | Dry | Yes | Yes | Yes | Yes | Yes | Yes | Yes |
| 45 | (Marvi et al., 2014) | Sand | Animal, Robot | Sidewinding | Dry | No | Yes | Yes | Yes | Yes | Yes | No |
| 46 | (Mazouchova et al., 2010) | Sand | Animal | Crawl | Dry | No | Yes | Yes | Yes | Yes | Yes | No |
| 47 | (Mazouchova et al., 2013) | Sand | Animal, Robot | Crawl | Dry | No | Yes | Yes | Yes | Yes | No | No |
| 48 | (McInroe et al., 2016) | Sand | Animal, Robot | Crutch with tail use | Dry | No | Yes | Yes | Yes | Yes | Yes | No |
| 49 | (Naclerio et al., 2021) | Sand | Robot | Vertical intrusion | Dry | No | Yes | Yes | Yes | Yes | No | No |
| 50 | (Ortiz et al., 2019) | Sand | Probe | Drag | Dry | No | No | Yes | Yes | Yes | No | No |
| 51 | (Pravin et al., 2021) | Sand | Probe | Vertical intrusion | Dry | No | Yes | Yes | No | No | No | No |
| 52 | (Qian and Goldman, 2015a) | Sand | Robot | Run | Dry | Yes | Yes | No | Yes | Yes | Yes | No |
| 53 | (Qian and Goldman, 2015b) | Sand | Robot | Run | Dry | Yes | Yes | No | Yes | Yes | No | No |
| 54 | (Qian et al., 2012) | Sand | Robot | Walk | Dry | No | Yes | Yes | Yes | Yes | No | Yes |
| 55 | (Qian et al., 2015) | Sand | Animal, Robot | Walk | Dry | Yes | Yes | Yes | Yes | Yes | Yes | No |
| 56 | (Roberts and Koditschek, 2018) | Sand | Robot | Jump | Dry | No | No | Yes | Yes | Yes | No | No |
| 57 | (Schiebel et al., 2020) | Sand | Animal, Robot | Lateral undulation | Dry | No | Yes | Yes | Yes | Yes | Yes | No |
| 58 | (Sharpe et al., 2013) | Sand | Animal | Walk, burrow, swim | Dry | Yes | Yes | Yes | Yes | Yes | Yes | Yes |
| 59 | (Sharpe et al., 2015) | Sand | Animal | Burrow, swim | Dry, wet | Yes | Yes | Yes | Yes | Yes | Yes | Yes |
| 60 | (Shrivastava et al., 2020) | Sand | Robot | Wheeled | Dry, wet | No | Yes | Yes | Yes | Yes | No | No |
| 61 | (Tang et al., 2024) | Sand | Robot | Burrow | Dry | No | No | Yes | Yes | Yes | No | No |
| 62 | (Tao et al., 2020) | Sand | Animal, Robot | Burrow | Dry | No | No | No | Yes | Yes | No | No |
| 63 | (Thoesen et al., 2018) | Sand | Probe | Screw-powered propulsion | Dry | No | No | Yes | Yes | Yes | No | No |

| No. | Reference | Substrate | Locomotor | Locomotor mode/movement type | Wetness level | Varied ϕ | Controlled ϕ | Studied locomotor forces | Studied kinematics | Studied behavior | Studied motor control | Studied gait transition |
|---|---|---|---|---|---|---|---|---|---|---|---|---|
| 64 | (Thoesen et al., 2019a) | Sand | Robot | Screw-powered propulsion | Dry | No | No | No | Yes | Yes | No | No |
| 65 | (Thoesen et al., 2019b) | Sand | Probe | Screw-powered propulsion | Dry | No | No | Yes | No | Yes | No | No |
| 66 | (Thoesen et al., 2020a) | Sand | Robot | Wheeled | Dry | No | No | No | No | Yes | No | No |
| 67 | (Treers et al., 2021) | Sand | Probe | Drag | Dry | No | No | Yes | No | Yes | No | No |
| 68 | (Treers et al., 2022) | Sand | Animal, Robot | Burrow | Dry, saturated | No | No | Yes | Yes | Yes | No | No |
| 69 | (Winter et al., 2012) | Sand | Animal | Burrow | Saturated | No | No | Yes | Yes | Yes | No | No |
| 70 | (Xiong et al., 2017) | Sand | Robot | Walk | Dry | No | No | Yes | Yes | Yes | Yes | No |
| 71 | (Yu et al., 2024) | Sand | Probe | Wheeled | Dry | No | No | Yes | No | Yes | No | No |
| 72 | (Zhang et al., 2013a) | Sand | Robot | Run | Dry | No | Yes | Yes | Yes | Yes | No | No |
| 73 | (Zhang et al., 2017) | Sand | Animal | Run, walk | Dry | No | No | Yes | No | Yes | No | No |
| 74 | (Zhang et al., 2018) | Sand | Animal | Run, walk | Dry | No | No | No | Yes | Yes | Yes | No |
| 75 | (Kerimoglu et al., 2025) | Powder | Probe | Vertical intrusion | Dry | No | Yes | Yes | No | Yes | No | No |
| 76 | (Schiebel et al., 2022) | Homogeneous substrate | Robot | Run | Dry | No | No | No | Yes | Yes | No | No |
| 77 | (Thoesen et al., 2020b) | Lunar analog | Robot | Screw-powered propulsion | Dry | No | No | No | Yes | Yes | No | No |
| 78 | (Henmi and Itani, 2014) | Mud-sand mixture | Animal | Burrow utilization | Saturated | Yes | No | No | No | Yes | No | No |
| 79 | (Liu et al., 2023) | Mud-sand mixture | Robot | Crutch | Semi-solid | Yes | Yes | Yes | Yes | Yes | Yes | No |
| 80 | (Francoeur and Dorgan, 2014) | Sand, Viscoelastic solid | Animal | Burrow | Semi-solid, wet | N/A (No for sand) | N/A for gelatin (No for sand) | No | Yes | Yes | Yes | No |
| 81 | (Murphy and Dorgan, 2011) | Viscoelastic solid | Animal | Burrow | Semi-solid | N/A | N/A | Yes | Yes | Yes | No | No |
| 82 | (Che and Dorgan, 2010a) | Viscoelastic solid | Animal | Burrow | Semi-solid | N/A | N/A | No | Yes | Yes | No | No |
| 83 | (Che and Dorgan, 2010b) | Viscoelastic solid | Animal | Burrow | Semi-solid | N/A | N/A | No | Yes | Yes | No | No |
| 84 | (Dorgan et al., 2007) | Viscoelastic solid | Animal | Burrow | Semi-solid | N/A | N/A | Yes | Yes | Yes | No | No |

| No. | Reference | Substrate | Locomotor | Locomotor mode/movement type | Wetness level | Varied ϕ | Controlled ϕ | Studied locomotor forces | Studied kinematics | Studied behavior | Studied motor control | Studied gait transition |
|---|---|---|---|---|---|---|---|---|---|---|---|---|
| 85 | (Dorgan et al., 2013) | Viscoelastic solid | Animal | Burrow, swim | Semi-solid, Fragmented semi-solid | N/A | N/A | Yes | Yes | Yes | Yes | No |
| 86 | (Wang et al., 2013) | Viscoelastic solid | Animal | Crutch, swim | Semi-solid | N/A | N/A | No | Yes | Yes | Yes | Yes |
| 87 | (Horner and Jayne, 2008) | Viscous fluid | Animal | Swim | Fluid, semi-fluid | Yes | Yes | No | Yes | Yes | Yes | No |
| 88 | (Lutek and Standen, 2019) | Viscous fluid | Animal | Swim | Fluid, semi-fluid | Yes | Yes | No | Yes | Yes | Yes | No |
| 89 | (Lutek and Standen, 2021) | Viscous fluid | Animal | Swim | Fluid, semi-fluid | Yes | Yes | No | Yes | Yes | Yes | No |
| 90 | (Bagheri et al., 2020) | Sand, Mud | Animal | Walk, run | Dry, saturated | No | No | No | Yes | Yes | Yes | Yes |
| 91 | (Falkingham and Horner, 2016) | Sand, Mud | Animal | Walk | Wet, semi-solid | No | Yes | No | Yes | Yes | Yes | No |
| 92 | (Liang et al., 2012) | Sand, Mud | Robot | Walk | Dry, wet, Semi-solid, semi-fluid | Yes | Yes | Yes | Yes | Yes | No | No |
| 93 | (Naylor and Kawano, 2022) | Sand, Mud | Animal | Crutch | Semi-solid, dry | No | Yes for mud (No for sand) | No | Yes | Yes | Yes | No |
| 94 | (Standen et al., 2016) | Sand, Mud, Pebbles | Animal | Walk | Wet, semi-solid | No | No | No | Yes | Yes | Yes | No |
| 95 | (Tomie et al., 2017) | Sand, Mud, Pebbles, Gravel, Cobble | Animal | Burrow | Saturated | No | No | No | Yes | Yes | No | No |
| 96 | (Zhang et al., 2013b) | Sand, Mud | Robot | Walk | Dry, wet, semi-solid, semi-fluid | Yes | Yes | Yes | Yes | Yes | No | No |
| 97 | (Davenport and Matin, 1990) | Mud | Animal | Walk | Firm | No | No | No | No | Yes | No | No |
| 98 | (Du et al., 2016) | Mud | Animal | Walk | Semi-solid | No | No | No | No | Yes | No | No |
| 99 | (Godon et al., 2024) | Mud | Robot | Walk | Semi-solid | No | Yes | Yes | Yes | Yes | Yes | No |
| 100 | (Harris, 1960) | Mud | Animal | Skip, crutch | None | No | No | No | Yes | No | No | No |
| 101 | (Horner and Jayne, 2014) | Mud | Animal | Walk | Wet | No | Yes | No | Yes | Yes | Yes | No |

| No. | Reference | Substrate | Locomotor | Locomotor mode/movement type | Wetness level | Varied ϕ | Controlled ϕ | Studied locomotor forces | Studied kinematics | Studied behavior | Studied motor control | Studied gait transition |
|---|---|---|---|---|---|---|---|---|---|---|---|---|
| 102 | (Ren et al., 2013) | Mud | Robot | Walk | Solid, semi-solid, semi-fluid, fluid | Yes | Yes | No | Yes | Yes | No | No |
| 103 | (Riskowski and DeShazer, 1976) | Mud | Animal | Walk | Semi-solid | No | No | No | No | Yes | No | No |
| 104 | (Zhang et al., 2016) | Mud | Robot | Walk | Solid, semi-solid, semi-fluid, fluid | Yes | Yes | No | Yes | Yes | Yes | Yes |
| 105 | (Dorgan et al., 2016) | Mud, Viscoelastic solid | Animal | Burrow | Semi-solid | N/A (No for mud) | N/A for gelatin (No for mud) | No | Yes | Yes | Yes | No |
| 106 | (Kuznetsov, 2022) | Mud | Animal | Walk | Solid, semi-solid | Yes | No | No | No | Yes | No | No |

The methods for controlling dry and wet sandy substrates cannot be used to prepare muddy substrates. Dry sand can be controlled to the desired volume fraction using an air fluidized bed which blows air through grains to fluidize them into a loosely packed state, followed by air pulses or shaking to compact the grains (Li et al., 2009; Maladen et al., 2009), but this does not work for solid–water mixtures. Wet sand can be prepared using a shaker that makes moist grains fall through a sieve to deposit to a uniform state of a controlled wetness (Sharpe et al., 2015), but this cannot be applied to sticky muddy substrates which can clog the sieve.

In addition, for muddy substrates (and unsaturated wet sand), water evaporation increases solid volume friction and reduces wetness, potentially causing it to enter another regime (Fig. 1B). Thus, it is useful to seal the prepared substrate when not in use and track its yield strength drift.

Furthermore, we speculate that mud is much weaker than other flowable substrates like dry sand. In our preliminary experiments, commercially available penetrometers for characterizing substrate yield strength (by measuring vertical resistive force on a probe pushed into the substrate to a fixed depth) widely used for sand and soil are not sensitive enough for muddy substrates except when around or above the fracture limit. We need new tools for characterizing the yield strength of muddy substrates over the broad range of naturally occurring conditions. This also suggested that muddy substrates can be much more challenging as animals and robots can easily get bogged down.

Here, we developed low-cost tools and methods to precisely control, systematically vary, accurately characterize, and temporally track the drift of the yield strength of muddy substrates with intermediate wetness that can go through solid–fluid transitions (regime d in Fig. 1B). To test and demonstrate their usefulness, we chose clay mud, a mixture of fine clay and water with no coarse grains and of intermediate wetness with solid volume fraction $\phi$ between the settling $\phi_s$ and fracture $\phi_f$ limits (Fig. 1B, green right boundary), because of its relative simplicity. Natural muddy substrates can consist of both fine clay and coarse grains, and their yield strength is a function of both $\phi$ and $\eta$ (Coussot, 2017). But for clay mud consisting of a given kind of fine clay, its yield strength simply increases monotonically with $\phi$ (by up to several orders of magnitude) (Coussot, 2017). Despite this relative simplicity, clay mud display qualitatively similar rheology as mud with both fine clay and coarse grains in the solid–fluid transition regime (Fig. 1B, green right boundary vs. green band) (Coussot, 2017). So, we expect that these tools and methods can be applied to natural muddy substrates with the solid–fluid transition regime (regime d in Fig. 1B).

## 2. Materials and Methods

### 2.1. Clay selection

To prepare clay mud, we chose to use Kaolin clay, which has a relatively low cost (~$0.4–4 per kg). Kaolin clay has been extensively used in previous mud studies (for a review, see (Coussot, 2017)), so our data can be directly compared with previous results. In preliminary experiments (Sec. 2.14, Fig. S1; Sec. 3.3–3.4, Fig. 3C–E), we used Edgar Plastic Kaolin (EPK) clay (Edgar Minerals, USA). Later, when Edgar Plastic Kaolin clay became unavailable for a long time, we used Georgia Kaolin (China Clay, Old Hickory Clay Company, Florida, USA) in all subsequent experiments (comparison of stationary automatic and portable manual penetrometer, Supplementary Information Fig. 5A–B, tracking large drift in mud strength for long experiments, Sec. 3.5, Fig. 4). We verified that Edgar Plastic Kaolin clay (Fig. 1C) and Georgia Kaolin clay (Fig. 1D) behave qualitatively similar, though the quantitative results did differ. Our Edgar Plastic Kaolin clay powder had larger in particle size than to Georgia Kaolin clay powder. Such qualitatively similar behavior with quantitative difference between mud with different Kaolin clay has been noted (Coussot, 2017). Thus, it is best to use the exact same kind of clay whenever possible to conduct repeatable experiments.

### 2.2. Calculation of clay-to-water ratio for a given wetness

Following the more common practice in the flow rheology literature (Coussot, 2017), we used solid volume fraction ϕ as a measure for solid concentration, which decreases with increasing wetness (water concentration). ϕ is defined as the ratio between the volume of solid particles $V_s$ to the total volume of the solid–water mixture, which also contains water volume $V_w$:

$$\phi = \frac{V_s}{V_s + V_w} \quad (1)$$

One may also use solid weight fraction $w$, defined as ratio between the weight of solid particles $w_s$ to the total weight of the mixture, which also contains water weight $w_w$:

$$w = \frac{w_s}{w_s + w_w} \quad (2)$$

ϕ and $w$ can be converted into each other, knowing the solid particle density $\rho_s$ ($2.64 \times 10^3$ kg/m$^3$ for Kaolin clay) and water density $\rho_w$ ($1 \times 10^3$ kg/m$^3$):

$$w = \frac{1}{1 + \frac{\rho_w}{\rho_s}\left(\frac{1}{\phi\%} - 1\right)} \quad (3)$$

To prepare clay mud of a desired ϕ, we calculated and precisely measured (using a 220 LB high accuracy electronic digital refrigerant charging weight scale, 220 pounds maximum capacity, 5 g readout accuracy, Wale&Morn Store) the weight of kaolin clay $w_s$ needed for a given weight of water $w_w$ using Eqns. 2 and 3.

### 2.3. Automated mud mixer

To prepare a large quantity of mud uniformly in relatively a short amount of time, we developed a low-cost automated mud mixer (Fig. 2A). This system is useful because our preliminary experiments showed poor uniformity even across small batches of mud mixed by hand or even using an electric drill with a high viscosity helical mixing paddle (0.23 m diameter, 0.61 m length, Model MG 235, CS Unitec, Norwalk, CT, USA). Our system improved over these by maintaining a constant, slow rate of clay being fed into water for uniform mixing. This system consisted of an electric stand mixer (ACA, model A86208, JEAHII, Guangzhou KeYue Technology Co. Ltd., Guangzhou, China), a custom-made dipper made of cardboard sheets, and a 3-D printed rotating cam (Fig. 2A). The dipper holding the clay powder sat on an inclined T-slotted framing (McMaster-Carr, Princeton, NJ, USA) (Fig. 2A). We manually increased the dipper incline as the amount of clay powder reduced. A DC motor (uxcell, 12 V, 125 RPM, with Worm Gear, Hong Kong, China) (Fig. 2A) drove the cam to raise and lower the incline to repeatedly shake a small amount of clay powder into the bowl of water every 0.5 seconds. This is important because a large amount of clay powder poured into water tended to form clumps and did not mix uniformly with water. We operated mixer motor between 3$^{rd}$ (88 RPM) and 5$^{th}$ (100 RPM) speed setting, slowly enough to mix uniformly. With the automatic mixer, depending on the solid volume fraction $\phi$, it took 10–20 minutes to make one batch of clay mud with 1.5 kg of water. The total cost of the mixer was ~$166.

To fill larger testbeds (Fig. 2C), multiple batches of mud were prepared and combined. For example, to fill a testbed of 1.02 m × 0.51 m × 0.16 m in volume with mud of 49.3 kg total weight of $\phi$ = 34% used in our companion animal study (Ramesh et al., submitted), 14 batches (each 3.5 kg) were needed, taking a total of 5 hours over ~2.5 days (not full-time operation). After mixing, each batch of mud was transferred from the mixer into the large container and mixed manually with the existing mud for at least 10 minutes using a metal spatula with beveled edges (Homi Styles) to ensure consistency, and a sealing method (Sec. 2.7) was applied immediately to minimize water evaporation and drift in solid volume fraction $\phi$.

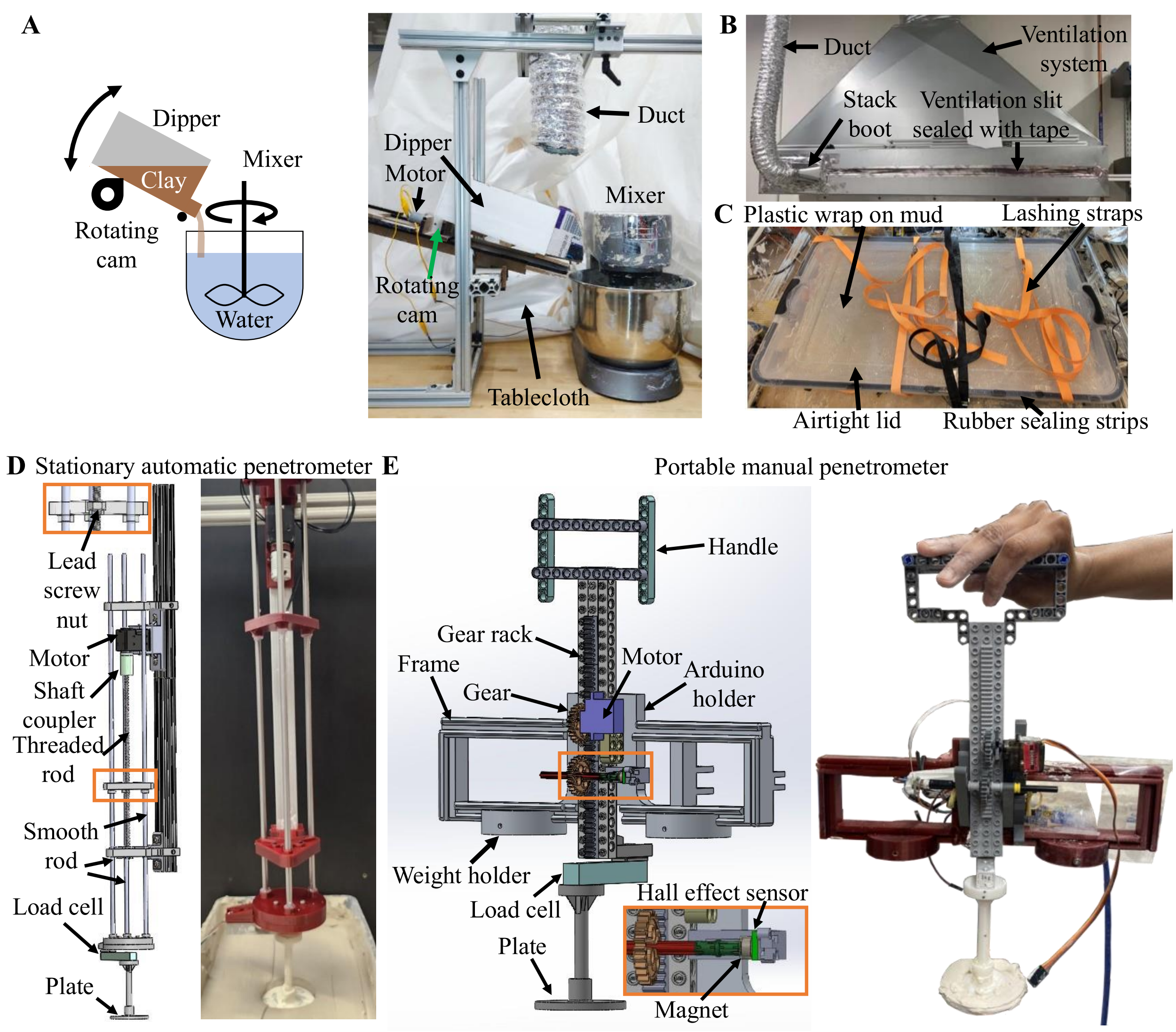

**Fig. 2. Mud preparation, storage, and characterization tools. (A)** Automated mud mixer. **(B)** Ventilation system for safe clay handling and mixing. **(C)** Sealed container to minimize water evaporation between use. **(D, E)** Stationary automatic and portable manual penetrometer.

## 2.4. Safety during mud preparation

Safety measures are important to ensure user health during mud preparation. Clay particles are so small that they can become airborne and pose health risks when inhaled. Kaolin clay can contain a small amount of crystalline silica (0.5% by weight), which can cause various diseases (Occupational Safety and Health Administration). To mitigate these risks, a ventilation system is crucial. Our lab has a ventilation system installed to remove small particles by sucking the air rapidly when turned on (Fig. 2B). In general, any powerful air ventilation system with small particle filter (e.g., HEPA) is useful for this purpose. A simple custom duct can be added to maximize the efficiency of dust removal near the mixing apparatus. In our case, we used a flexible foil duct tube (0.1 m diameter, 7.62 m long, Everbilt, Home Depot, Atlanta, GA, USA) routed from the ventilation system using a stack boot (Master flow, Home Depot, Atlanta, GA, USA) to directly above the mixer (Fig. 2A–B). We also surrounded the automated mixer with a plastic tablecloth (Fig. 2A, Exquisite by Crown Display, West Pittston, PA, USA) to contain remaining clay dust within a small space and prevent exposure to the entire room. For maximal dust removal, the vent was turned on before the start of the mud preparation and was turned off a 2–3 hours after mud preparation. To further minimize potential inhalation of crystalline silica, the user should wear N95 masks and long sleeve gloves (VGO Gloves, City of Industry, CA, USA) when removing the clay powder from its container, weighing it, and pouring it into the dipper, as well as wash the clothes worn after mud mixing with water at the end of the day. The Health Safety and Environment Homewood Safety office at Johns Hopkins University performed crystalline silica and particulate sampling tests during the use of the ventilation system at the time of mud mixing. For the respirable crystalline monitoring test (personal monitoring for the individual preparing the mud), the exposure of crystalline silica and respirable dust, which included the time of ~ 5 hours and 1 hour during and after the mixing operation, respectively, were found to be less than 6.9 $\mu g/m^3$ and equal to 0.047 $mg/m^3$, respectively. The exposures including respirable dust exposure from the particle monitoring test (exposure estimation for person working in the same space) performed for same duration above 21 days after mixing operation (taken for a duration of ~ 6 hours) were well below the OSHA permissible exposure limit for respirable dust (5 $mg/m^3$). Similar tests should be carried out, if possible, when the system is set up elsewhere.

### 2.5. Determining wetness range of solid–fluid transition regime

To determine the wetness range of the solid–fluid transition regime (Fig. 1B, regime d) of our clay mud, we need to know the settling and fracturing limits $\phi_s$ and $\phi_f$. To determine the settling limit, we slowly increased $\phi$ in small increments (by 1–2% each time) and tested whether there is settling of mud for each increment. We started with fluid mud of a low $\phi$ (e.g., 5%, Fig. 1D). After sufficient mixing (e.g., 10 minutes using the automatic mixer), we allowed it to settle for 5 minutes. Then, we used a spatula to scoop up the bottom content and visually checked whether it contained significantly more solid particles than the fluid mud above. If so, we then repeated the same procedures with fluid mud of higher $\phi$ (in increments of 1% to 5%). The $\phi$ at which there was no obvious settling of solid particles after 5 minutes then provided a good approximate (lower bound) of the settling limit. More accurate measurement of the settling limit can be achieved by waiting for a much longer time, e.g., 24 hours (Coussot, 2017).

To determine the fracture limit, we increased $\phi$ in small increments (3–5%) and tested for fracturing of mud after each increment. We started with mud with a $\phi$ in the solid–fluid transition regime (e.g., 20%; Fig. 1D). After sufficient mixing (e.g., 20 minutes using the automatic mixer), we flattened the mud surface using a metal spatula with beveled edges. We then intruded downward into the mud using a probe (e.g., a circular plate) and/or dragged an object (e.g., a metal spatula with beveled edges) horizontally across the surface. (Mud made from Georgia Kaolin clay, with finer particles than Edgar Plastic Kaolin clay, was harder to fracture using horizontal dragging, so vertical penetration was more reliable.) If the mud was readily fractured (e.g., formation of small cracks from the intruder from intrusion or dragging which did not close up or reset itself after 2 minutes) during intrusion and/or dragging, its $\phi$ had exceeded the fracturing limit (Coussot, 2017). More accurate measurement of the fracture limit can be achieved by waiting for a much longer time, e.g., 1 hour.

### 2.6. Clay and solid volume fraction used for each test

As mentioned, we used two kinds of Kaolin clay, Edgar Plastic Kaolin clay and Georgia Kaolin clay. For mud spatial uniformity and short-term temporal drift tests (Sec. 2.14), we used Edgar Plastic Kaolin clay and made clay mud of $\phi$ = 14%, 27%, 34%, and 39%. For testing boundary effect during mud characterization (Sec. 2.8), we used Edgar Plastic Kaolin clay and made clay mud of $\phi$ = 34% and 39%. Most of these fell within the solid–fluid transition regime between the settling and fracture limits ($\phi_s < \phi <$ $\phi_f$, Fig. 1C), with $\phi$ = 14% being around the settling limit $\phi_s$. For characterizing dependence on mud strength wetness test (Sec. 2.15) and long-term mud strength drift tests (Sec. 2.16), we used Georgia Kaolin clay and made clay mud of $\phi$ = 27%, 34%, 39%, and 42%, most of which fell within the solid–fluid transition regime between the settling and fracture limits ($\phi_s < \phi < \phi_f$, Fig. 1D) with 42% above fracture limit ($\phi > \phi_f$).

**2.7. Mud container sealing**

For more repeatable experiments and better comparison of results over a long time, we must minimize drift in $\phi$ from water evaporation. We used a container with an airtight lid (1.02 m × 0.51 m × 0.16 m, HOMZ, Chicago, IL, USA) (Fig. 2C). To further minimize water evaporating from mud surface into the remaining space under the lid, we attached a plastic wrap (used for food storage, Saran Wrap, Midland, MI, USA) to the mud surface. Finally, we used rubber sealing strips (CloudBuyer) and latching straps (ACE-Lashing Straps, Acelane) to tight down the lid (Fig. 2C). We tested how well these methods worked by tracking the drift in mud strength over multiple days (see results in Sec. 2.16).

These sealing methods were used whenever mud was not being used for experiments to minimize drift in mud properties.

**2.8. Mud container size to minimize boundary effect**

It is important to ensure that the mud container is sufficiently large so that the force-measuring probe for mud characterization (Sec. 2.9–2.13) does not move near the bottom and sidewalls of mud containers to minimize boundary effect. When this happens, the measured force become larger than it would be in an infinitely large container (Coussot, 2017) (see Fig. S1 for an example). For the large container that we used (1.02 m length × 0.51 m width × 0.16 m depth, HOMZ, Chicago, IL, USA), boundary effect from the sidewalls can be easily avoided. To minimize boundary effect from the bottom, we filled the container with mud to a height of 0.12 m (3/4 of the container height). We performed penetration experiments (following procedures in Sec. 2.10–2.11) and found no significant boundary effect for Edgar Plastic Kaolin clay mud of $\phi$ = 34% and $\phi$ = 39% as long as the container was filled to > 10 cm (Fig. S1B, solid), whereas boundary effect was observed when the container was filled to a lower height (~ 3  cm, Fig. S1B, dashed).

### 2.9. Custom-made penetrometers to characterize mud strength

A common, simple way to characterize the yield strength of a flowable substrate (either sandy or muddy) is to measure how strongly it resists penetration of an intruder of a given size. For example, the upward force (lift) on a horizontal disc of a given area pushed downward into the substrate increases as the substrate becomes stronger (Li et al., 2009; Li et al., 2013). This upward force is proportional to the intruder area. This test can resolve substrate property variation on spatial scales larger than intruder size by moving the test location. Resolving finer spatial variation needs other techniques such as x-ray imaging (Sharpe et al., 2015).

In early experiments, we tested three commercial penetrometers, including a Geotester pocket dial penetrometer ($355, Model HM-502, with a 0.025 m diameter probe penetrating to a 0.0064 m depth, Gilson Company Inc, Ohio, USA), a pocket penetrometer ($57, Model 59032, loading piston with 0.0064 m diameter, AMS Inc, American Galls, ID, USA), and a pocket soil penetrometer ($80, Model 5DPJ8, Humboldt Mfg. Co., Elgin, IL, USA). Only the most expensive of the three (Geotester pocket penetrometer) was sensitive enough to measure vertical force of the 0.025 m diameter disc probe penetrated to a 0.0064 m depth in the stronger Edgar Plastic Kaolin clay mud of $\phi$ = 39% right below the fracture limit, and even this penetrometer was not sensitive enough for most of the range of $\phi$ of the solid–fluid transition regime. This was likely because these commercial penetrometers were designed more for application in stronger substrates such as soil and dry sand, which are about 1–2 orders of magnitude stronger than mud of most $\phi$ in the solid–fluid transition regime (see results in Fig. 5L). In addition, commercial penetrometers are designed to penetrate to a fixed depth but cannot measure how substrate force changes with depth during probe intrusion and extraction, which is important for understanding the complex, spatiotemporally changing locomotor challenges by the substrate (Fig. 3A, 5A–B).

Therefore, to better characterize the yield strength of muddy substrates in the solid–fluid transition regime (Fig. 1B, d), we developed two custom penetrometers that are sufficiently sensitive and can measure substrate force as a function of depth. The stationary automatic penetrometer (Fig. 2D) is heavy and fixed to a table and automatically actuated to penetrate the probe at a constant speed to a prescribed maximal depth as the vertical force is measured. The portable manual penetrometer (Fig. 2E) is small and lightweight and can be easily moved to measure at different locations, though its manual operation leads to less precise control of the penetration speed and depth.

### 2.10. Stationary automatic penetrometer

We developed a stationary automatic penetrometer to characterize mud by measuring vertical force as a function of depth (Fig. 2D). This penetrometer was 1.61 kg in weight, 70 cm in height, 5.5 cm in width, and 13 cm in thickness (into the plane in Fig. 2D). It can move the intruder vertically up to 18 cm and can measure forces up to 49 N, with a position resolution of 0.0011 mm (estimated from a 0.25° mechanical backlash, which exceeded the 0.088° angular resolution of the servo motor) and a force resolution of 10 μN. The total cost of this stationary penetrometer was ~$300.

The stationary penetrometer used a custom linear translation system to push an intruder disc into a substrate at a controllable, constant speed, a force sensor to measure the vertical substrate resistive forces on the disc, and a position sensor to measure vertical displacement. This penetrometer was secured to a frame made up of T-slotted framing (McMaster-Carr, Princeton, NJ, USA) mounted on a table, with care taken to ensure that the intruder disc was perpendicular to the mud surface. The penetrometer used a Dynamixel motor (XM430-W350-R, ROBOTIS, Seoul, Korea) to move the intruder vertically (Fig. 2D). The motor rotation was translated to linear displacement using a threaded rod (12” lead screw, 1/4”-16 thread size, McMaster-Carr, Princeton, NJ, USA). This threaded rod was connected to the motor shaft using a 3D-printed part and shaft coupler (uxcell, Hong Kong, China) and supported by three smooth rods (24” linear motion shaft, 1/4” diameter, McMaster-Carr, Princeton, NJ, USA) (Fig. 2D).

Force was measured by a single-axis force sensor (strain gauge load cell, 5 kg, S18X4, ShangHJ) installed between the intruder and the linear stage with its axis aligned with the vertical direction (Fig. 2D). The force data were collected using an Arduino nano (Arduino.cc, Monza, Italy) via a HX711 ADC module (ShangHJ), which provided highly precise digital readings of the force data. We used MATLAB to control the motor and collect the load cell reading and motor's present position reading to obtain force as a function of displacement (from which penetration depth was determined, see Sec. 2.11). The disc should be cleaned before every use so that no mud residue biases the data. The force sensor should be zeroed at the beginning of each trial.

In our experiments using the stationary automatic penetrometer, the intruder was a 3-D printed disc with a thickness of 0.5 cm. It was moved by a maximal depth of 4.96 cm into and out of the mud at 0.31 cm/s. For experiments with Georgia Kaolin clay mud, the radius of the disc was 3 cm for lower $\phi$ of 20%, 25%, 27%, 34%, 39%, and 41%, but for the higher $\phi$ = 42% and for dry mud it was reduced to 1 cm because the force on a 3 cm radius disc would exceed the sensor limit. For experiments with Edgar Plastic Kaolin clay mud, the radius of the disc was 3 cm for lower $\phi$ = 14% and 27% but for higher $\phi$ = 34% and 39%, it was reduced to 1 cm. Whenever comparing force data between these two cases, the measured force on the 1 cm radius disc was scaled to that expected for a 3 cm radius disc, considering that vertical penetration force on a horizontal intruder scales with intruder area.

### 2.11. Data processing for stationary automatic penetrometer

To illustrate how to use raw data from the our custom penetrometers to calculate mud yield strength, we collected data for a few representative volume fractions, $\phi$ = 20%, 25%, 27%, 34%, and 39% in the solid–fluid transition regime (Fig. 1B, regime d) and $\phi$ = 41% and 42% in the fractured solid mud regime (Fig. 1B, regime f) on Georgia Kaolin clay mud, as well as for $\phi$ = 14%, 27%, 34%, and 39% in the solid–fluid transition regime (Fig. 1B, regime d) on Edgar Plastic Kaolin clay mud, and processed them as follows. The disc position (black, Fig. S2A) and force (black, Fig. S2B) data were synchronized using time stamps recorded in both data. Position data were then interpolated to have the same sampling frequency as force data. The load cell readings had a small cyclic noise due to the small cyclic movement along the threaded rod. To remove this, we filtered force data using a band stop filter with a stopband frequency range of 0.05 Hz to 15 Hz. For each trial, the disc depth was obtained by offsetting the disc position to start from 0 when the intruder first contacted the mud surface, which was determined as when the force exceeded a small threshold of 0.03 N (red, Fig. S2A–B). We visually examined the resulting force vs. depth data plots. If offsetting using the force threshold was not correct, we then offset the position manually. The synchronized force and depth data were then plotted as force vs. depth (Fig. S2C). We performed all data analysis in MATLAB.

### 2.12. Portable manual penetrometer

The portable manual penetrometer used the same basic mechanism as the stationary automatic penetrometer, but it can be easily moved anywhere to characterize substrate strength on a large animal or robot locomotion testbed. This penetrometer was 0.545 kg in weight, 44.5 cm in height, 26 cm in width, and 10.5 cm in thickness. It can move the intruder vertically up to 14 cm and can measure forces up to 49 N, with a position resolution of 0.11 mm (estimated from a 0.5° angular resolution of the Hall effect sensor) and a force resolution of 10 μN. The total cost of this portable penetrometer was ~$60.

This penetrometer also used the same force sensor, HX711 ADC module, and Arduino Nano as those in the stationary automatic penetrometer to measure forces. The difference was mainly in the rotation to linear translation mechanism, as follows. A handle made up of LEGO parts was added for holding the penetrometer when pushing the intruder disc into and pulling it out of mud (Fig. 2E, S3C). A LEGO gear rack and a LEGO gear converted the rotation applied by the experimenter into a linear motion (Fig. 2E). One end of the axle connected to the gear was attached with a magnet whose rotation was detected using a Hall effect sensor (AS5048B, AMS, Austria). This encoder measured vertical displacement (Fig. 2E). In an early prototype, we found that pushing the disc into the mud without resistance resulted in rapid penetration, especially on lower $\phi$, which caused the magnet to rotate too fast for the sensor to detect its rotation. To address this, resistance must be added to slow down penetration, for example by adding a geared rotary damper. Alternatively, we used a servo motor (MG90S, Dorhea) attached to a gear and placed over the gear rack to add resistance when powered (Fig. 2E, S3C).

In our experiments using the portable manual penetrometer to characterize dependence of mud strength on wetness (Sec. 2.15) and to track the drift of mud strength over many days (Sec. 2.16), we built a custom guide rail (T-slotted framing, McMaster-Carr, Princeton, NJ, USA) placed over a large testbed to move the penetrometer to different locations precisely over multiple tests over multiple days. The penetrometer must be used sufficiently far away from the walls of the container and the probe only pushed up to a depth not too close to the bottom of the container to avoid boundary effects (Fig. S1). To ensure that the disc was always perpendicular to the mud surface, we further added a 3-D printed frame with 0.1 kg weights (Zinc casted slotted mass set, Eisco LLC, Honeoye Falls, NY, USA) on both sides of the penetrometer (Fig. 2E, S3E) to help the penetrometer securely rest on the frame during operation. However, these additions are not strictly necessary for general use of the portable manual penetrometer.

### 2.13. Data processing for portable manual penetrometer

To illustrate how to use raw data from our custom penetrometers for characterizing mud yield strength, we collected data for a few representative volume fractions, $\phi$ = 20%, 25%, 27%, 34%, and 39% in the solid–fluid transition regime (Fig. 1B, regime d) and $\phi$ = 41% and 42% in the fractured solid mud regime (Fig. 1B, regime f) on Georgia Kaolin clay mud. We processed these data as follows. We used the force data (black, Fig. S2E) to find the initial contact with the mud by the disc during penetration which was when the force was more than a small threshold of 0.3 N. Linear displacement (black, Fig. S2D) and vertical force (black, Fig. S2E) data were cropped at initial mud contact. The force and depth data (red, Fig. S2D–E) were then plotted as force vs. depth (Fig. S2F). We performed all data analysis in MATLAB.

For characterizing the long-term drift of mud strength tests (Sec. 2.16), we divided the vertical force data into penetration and extraction phases by estimating the end of vertical force, start, and end of extraction force using the displacement and time data. The vertical force and displacement were first cropped to have the data up to 3 cm in depth during the penetration phase for comparing across multiple locations, trials, and several days. The force data during the penetration phase was then interpolated to average the data points across different trials.

We also tested and verified overall force vs. depth profile accuracy from the portable manual penetrometer by comparing its mud characterization with the mud characterization from the stationary automatic penetrometer for mud of different strengths (5A, B).

### 2.14. Experiments to characterize spatial uniformity and short-term temporal drift of mud strength

We performed experiments to assess the spatial uniformity of the prepared mud's yield strength and short-term (~ an hour) temporal drift of mud strength. These experiments were done using Edgar Plastic Kaolin clay mud of $\phi$ = 39% filled to 0.28 m height in a bucket of 0.3 m diameter and 0.37 m total height (5-gallon bucket, Home Depot, USA). These were done earlier in our study and we used the commercial Geotester pocket penetrometer that was sensitive enough for them. This penetrometer penetrated a 0.025 m diameter probe to a 0.0064 m fixed depth and measured the resulting upward force.

For spatial uniformity, we measured vertical force at 4 different locations, with 3 trials each. For temporal drift, we then repeated the above procedure every 10 minutes over the course of 1 hour, which is the typical duration of animal experiments in a day. The force was averaged across the 3 trials for spatial comparison across the 4 locations, and this trial average was further averaged across the 4 locations at each time interval for temporal comparison.

To further compare the spatial variation and temporal drift to the range over which we can vary mud strength using our methods, the force for $\phi$ = 39% was averaged both spatially and temporally, which was then compared with the average force for Edgar Plastic Kaolin clay mud of $\phi$ = 14%, 27%, and 34%, and 39%. These latter force data were obtained from mud characterization taken using the stationary automatic penetrometer, so the force was read at the 0.0064 m depth and scaled to the 0.025 m diameter disc area of the commercial penetrometer, considering that the horizontal intruder's vertical penetration force scales with intruder area.

### 2.15. Experiments to characterize dependence on mud strength wetness

To characterize the yield strength of the prepared clay mud, we collected data of vertical force vs. depth for Georgia Kaolin clay mud of $\phi$ = 20%, 25%, 27%, 34%, and 39% in the solid–fluid transition regime (Fig. 1B, regime d) and $\phi$ = 41% and 42% in the fractured solid mud regime (Fig. 1B, regime f) using both the portable manual and stationary automatic penetrometers. We chose to measure up to 4 cm depth because significantly deeper penetration would result in forces exceeding the range of the force sensors. This maximal depth of 4 cm was well beyond the sinkage much larger than fish sinkage (~0.3 cm) (Ramesh et al., submitted). The mud was filled to a 0.28 m height in a bucket of 0.3 m diameter and 0.37 m total height (5-gallon bucket, Home Depot, USA). We collected 3 trials for each $\phi$ and processed data following steps in Sec. 2.11 and Sec. 2.13. After each trial, we manually re-mixed and flattened the mud to reset the disturbed mud.

### 2.16. Experiments to characterize long-term drift of mud strength

To track the long-term temporal drift from water evaporation, we measured mud yield strength over the course of many days during animal experiments in our companion study (Ramesh et al., submitted). We used the portable manual penetrometer to perform penetration experiments for Georgia Kaolin clay mud with $\phi$ = 27%, 34%, and 39% in the solid–fluid transition regime (Fig. 1D) and $\phi$ = 42% in the fractured solid mud regime (Fig. 1D) over 27, 114, 106, and 24 days, respectively (Sec. 3.5, Fig. 4). We prepared the dry mud by leaving the mud of $\phi$ =42% open to dry over 1 week and collected data on Day 0 (first day of the experiments).

To quantify the spatial uniformity and short-term temporal drift of a large quantity of prepared mud, we measured mud strength at multiple locations of the testbed each hour over the course of animal experiments which lasted ~1 to ~7 hours. For mud with $\phi$ = 27%, 34%, and 42%, on the first day of animal experiments (Day 0), we did penetration tests at 18 locations with one trial each for the first three penetration tests (on locations where mud was undisturbed by the animal). On subsequent days, the number of locations tested were reduced to 4 as we found that the mud was quite uniform spatially. Occasionally, on days when we had a few animal trials that lasted less than an hour, we only did one penetration test over 18 locations with one trial each. On $\phi$ = 39%, we still performed penetration tests at 18 locations with one trial each for each day when the penetrometer was improved despite showing spatial uniformity to ensure repeatability in the mud characterization results. For dry mud, we characterized mud for one penetration test across 4 locations with 3 trials each.

To quantify the long-term temporal over many different days, the vertical force between 0 and 1 cm depth from the penetration phase was averaged across different trials and locations for each penetration test completed in a day to compare the vertical force as a function of depth between penetration tests in a day and different days for mud of each strength (Sec. 3.5, Fig. 4A–D). To quantify the long-term temporal drift, we averaged the vertical force for all penetration tests in each day and compared the force at 1 cm across all days to track the drift in mud strength for a longer duration of time spanning several days (Sec. 3.5, Fig. 4E–H).

## 3. Results and Discussion

### 3.1. Vertical force vs. depth in mud is complex

Vertical force from the substrate changed with intruder depth in complex ways during both intrusion and extraction (Figs. 3A, 5A, B). The force vs. depth profile was qualitatively somewhat similar for mud of different $\phi$. During intrusion when the intruder moved downward, vertical force was always upward (i.e., a lift) and increased monotonically with depth nonlinearly, initially rapidly but then more slowly and plateauing. This was unlike dry and wet sand where the vertical force increases linearly with depth (Fig. 5C–G). As the intruder stopped and stayed static, vertical force decreased substantially but remained upward (we checked that this was from the mud, not the load cell deforming). As soon as the intruder began moving upward, the upward force quickly diminished. As the intruder continued moving upward during extraction, vertical force became downward, because the cohesive mud sticking to the intruder pulled it downward. This was unlike dry and wet sand which have minimal vertical force during extraction due to lack of strong cohesion (Fig. 5C–G). For mud in the solid–fluid transition regime (e.g., Georgia Kaolin mud of $\phi$ = 20–41%, Fig. 3A, i–vi), this downward force even became larger (in magnitude) as the intruder became shallower. For the mud tested in the fractured solid regime (Georgia Kaolin of $\phi$ = 42%), the downward force first became larger then smaller as the intruder became shallower. For all $\phi$ = 20–42%, the downward force persisted even after the intruder was lifted off the mud surface and only diminished when all the mud sticking to the intruder eventually broke off (Fig. 3A, S4A–B).

**3.2. Mud yield strength increases exponentially with solid volume fraction**

We used the upward force at a fixed depth (0.64 cm, which is similar to the sinkage of mudskippers in our companion animal study (Ramesh et al., submitted)) in the initial rapid increase phase during intrusion as a measure of mud strength to compare across mud of different wetness. For Georgia Kaolin clay mud, as $\phi$ increased from $\phi$ = 20% (slightly above the settling limit in the solid–fluid transition regime) to $\phi$ = 42% (slightly above the fracture limit in the fractured solid regime) and, mud strength as measured by this vertical force increased exponentially by about 80 and (Fig. 3B). For Edgar Plastic Kaolin clay mud, from $\phi$ = 14% (around the settling limit) to $\phi$ = 39% (slightly below the fracture limit) in the solid–fluid transitions regime, mud strength as measured by this vertical force increased exponentially by about 950 times (Fig. 3E).

**3.3. Controlled mud is spatially uniform**

Our mud preparation methods achieved excellent spatial uniformity of mud strength (Fig. 3C). For example, for Edgar Plastic Kaolin clay mud with $\phi$ = 39%, the vertical force on a disc of 1.25 cm radius penetrated to 0.64 cm depth at each of the four locations tested was 14.1 ± 0.6 N, 14.9 ± 0.3 N, 13.4 ± 0.3 N, and 14.7 ± 1.3 N, respectively (Fig. 3C). The trial-to-trial variation measured by coefficient of variation (standard deviation divided by mean) was small, only 4%, 2%, 2%, and 9% at the four locations. The vertical force averaged from the means of all four locations was 14.3 ± 0.7 N, with a small variation of ~4%.

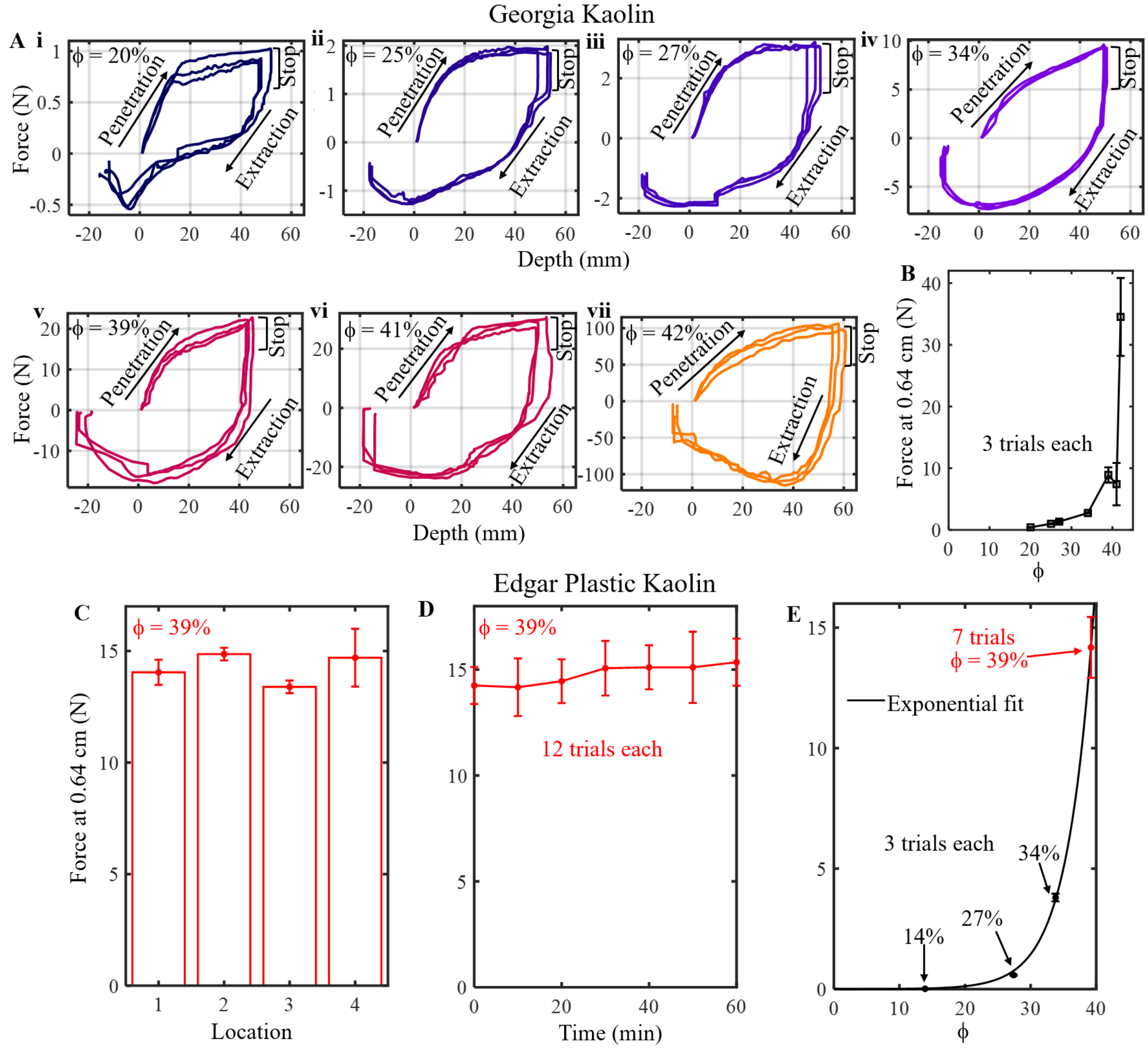

**Fig. 3. Vertical force during penetration to characterize mud yield strength and its dependence on wetness, spatial variation, and short-term temporal drift. (A)** Vertical force as a function of depth. Positive force is upward; negative force is downward. **(B)** Vertical force at a fixed depth (0.64 cm) vs. ϕ. Error bars in B show ± s.d. across trials. Both A, B used the portable manual penetrometer with a disc of a 3 cm radius and Georgia Kaolin clay mud, with 3 trials each. **(C, D)** Vertical force at a fixed depth (0.64 cm) at the four locations (C) and as a function of time over an hour (D). Error bars in C show ± s.d. of 3 trials at each location. Error bar in D show ± s.d. of all 12 trials across all locations. **(E)** Vertical force at 0.64 cm depth as a function of ϕ, with 3 trials each. Black error bars show ± s.d. of 3 trials for each ϕ. Red

error bar shows ± s.d. of the spatially averaged force across all time instances tested in D. Black curve is an exponential fit. C–E (red) used a commercial penetrometer with a disc of a 2.5 cm radius at 0.64 cm depth and Edgar Plastic Kaolin clay mud. E (black) used stationary automatic penetrometer.

### 3.4. Controlled mud strength drifts little within a short time

Within a short time (~1 hour) typical for completing one trial for all individuals in animal experiments in a day, there was only a small drift over time in mud yield strength measured by vertical force at a given depth during penetration (Fig. 3D). For example, for Edgar Plastic Kaolin clay mud with $\phi = 39\%$, after 60 minutes, the vertical force on a disc of a 1.25 cm radius penetrated to a 0.64 cm depth only increased by 8%, from 14.3 ± 0.9 N to 15.4 ± 1.1 N (Fig. 3D). Moreover, the overall spatiotemporal variation in yield strength for mud of a given $\phi$ was much smaller than the range of which yield strength can be varied by varying $\phi$ over the solid–fluid regime (e.g., a variation of ± 1.3 N vs. a 13.6 N range, red error bar at 39% vs. *y*-axis range of the data from 27% to 39% in Fig. 3E).

### 3.5. Strength drifts more over a long time but is much slower than without sealing

Our longer-term mud strength characterization experiment confirmed that the prepared mud strength changed relatively little (average of ~14%) within a day but drifted substantially (by ~2 times) over multiple days (Fig. 4).

On any given day, the vertical force at a given depth had only small spatial variations (Fig. 4A–D, same colored curves) of 15%, 16%, 13%, and 12% across 27, 114, 106, and 24 days for $\phi = 27\%$, 34%, 39%, and 42%, respectively (Fig. 4E–H, measured at 1 cm depth) and 14% across 1 day for dry mud (Fig. 4D, measured at 0.03 cm depth, as force far exceeded penetrometer range at larger depths).

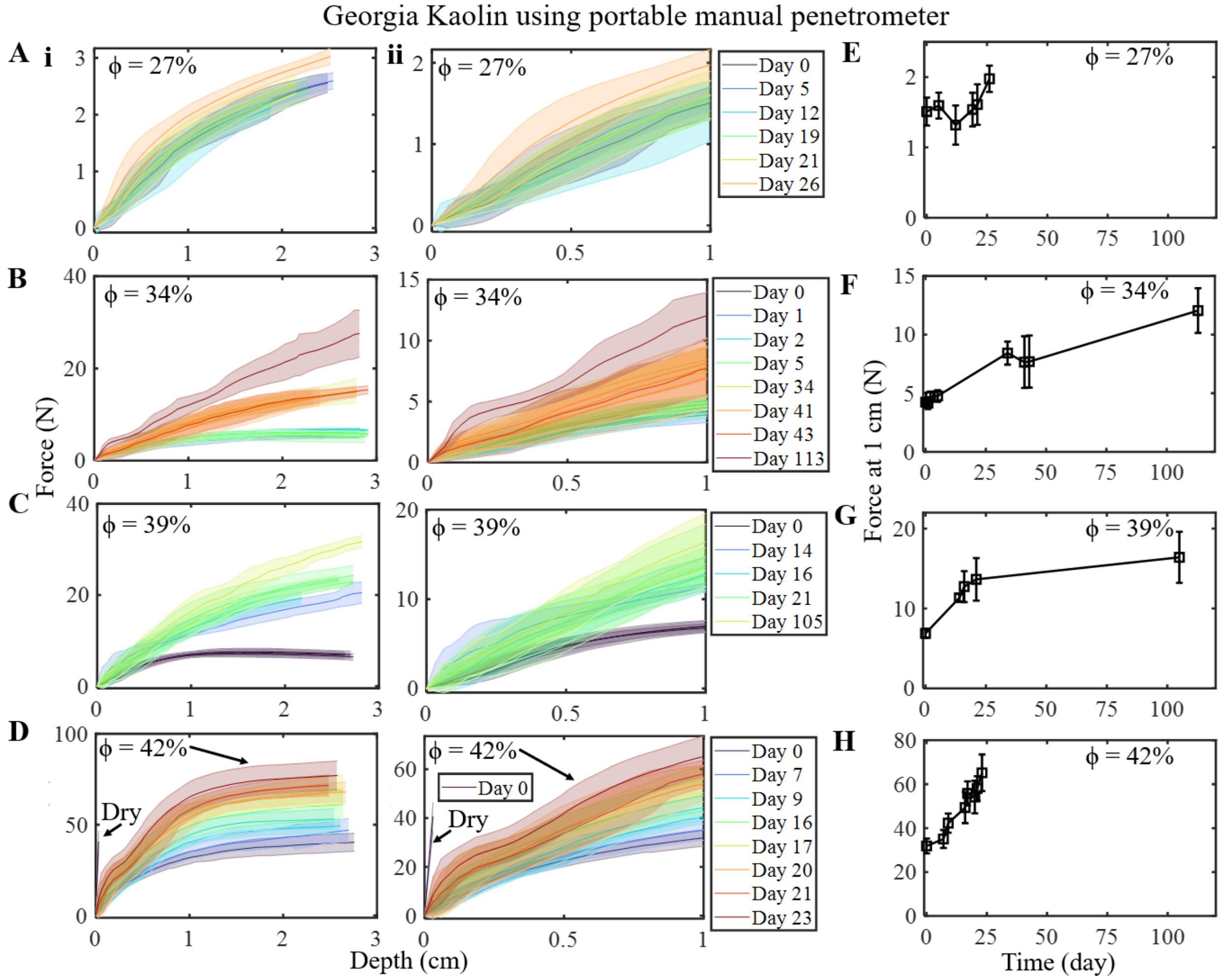

**Fig. 4. Tracking mud yield strength drift over many days. (A–D)** Vertical force as a function of depth during penetration. (i) Data of full range of depth measured. (ii) Data up to 1 cm depth. **(E–H)** Vertical force at 1 cm depth vs. the number of days elapsed from the first test. In all plots, error bars show ± s.d. of all trials on each day.

Over multiple days, the yield strength measured by vertical force at a given depth during penetration increased gradually but continuously (Fig. 4E–H), by an increase of 31% over 27 days for $\phi$ = 27%, an increase of 184% over 114 days for $\phi$ = 34%, an increase of 138% over 106 days for $\phi$ = 39%, and an increase of 104% over 24 days for $\phi$ = 42%. Nevertheless, our sealing methods ensured that this long-term drift is much smaller and slower than the rapid, large increase when leaving mud to dry in the open (e.g., increase by 133 times over just 7 days).

### 3.6. Mixer and sealing method apply to muddy substrates with both clay and grains

Although our methods were demonstrated for clay mud in the lab, they can be applied to or modified for natural muddy substrates in the solid–fluid transition regime (Fig. 1B, regime d), which have qualitatively similar rheology as clay mud (Coussot, 2017).

The automated mud mixer can be modified for mixing natural mud with both fine clay and coarse grains (regimes b–e, Fig. 1B) by using two rotating cam dippers to feed grains and clay, respectively, into water simultaneously at a prescribed ratio.

The sealing method should be used for such natural mud, as well as wet sand, to minimize water evaporation during storage to enhance experimental repeatability.

### 3.7. Mud is more challenging for locomotion than other flowable substrates

Comparing our mud characterization data with similar characterization using vertical penetration data from dry sand, wet sand, soil, and snow (Fig. 5C–K) demonstrated that mud in the solid–fluid transition regime is more challenging to move on than many other flowable substrates (Fig. 5).

First, the yield strength of wetter mud, measured by vertical force on an intruder of a given area at a given depth during penetration, can vary by 3 orders of magnitude with wetness, a much larger range than those of the other flowable substrates (usually up to than 1 order of magnitude) as their wetness or compaction changes.

In addition, mud strength in the solid–fluid transition regime is typically over an order of magnitude smaller than dry sand (even loosely packed) and snow, and it is two orders of magnitude smaller than wet sand and soil (Fig. 5L). Even as mud becomes drier (e.g., Georgia Kaolin clay mud of $\phi$ = 42% slightly above the fracture limit), it is still weaker than wet sand and soil and only comparable in strength to dry sand and snow (Fig. 5L). This means that the same animal, robot, or vehicle would sink one to two orders of magnitude more in wetter mud, and at least as much on drier mud. Only dry mud far above the fracture limit is stronger, as it is a fractured solid.

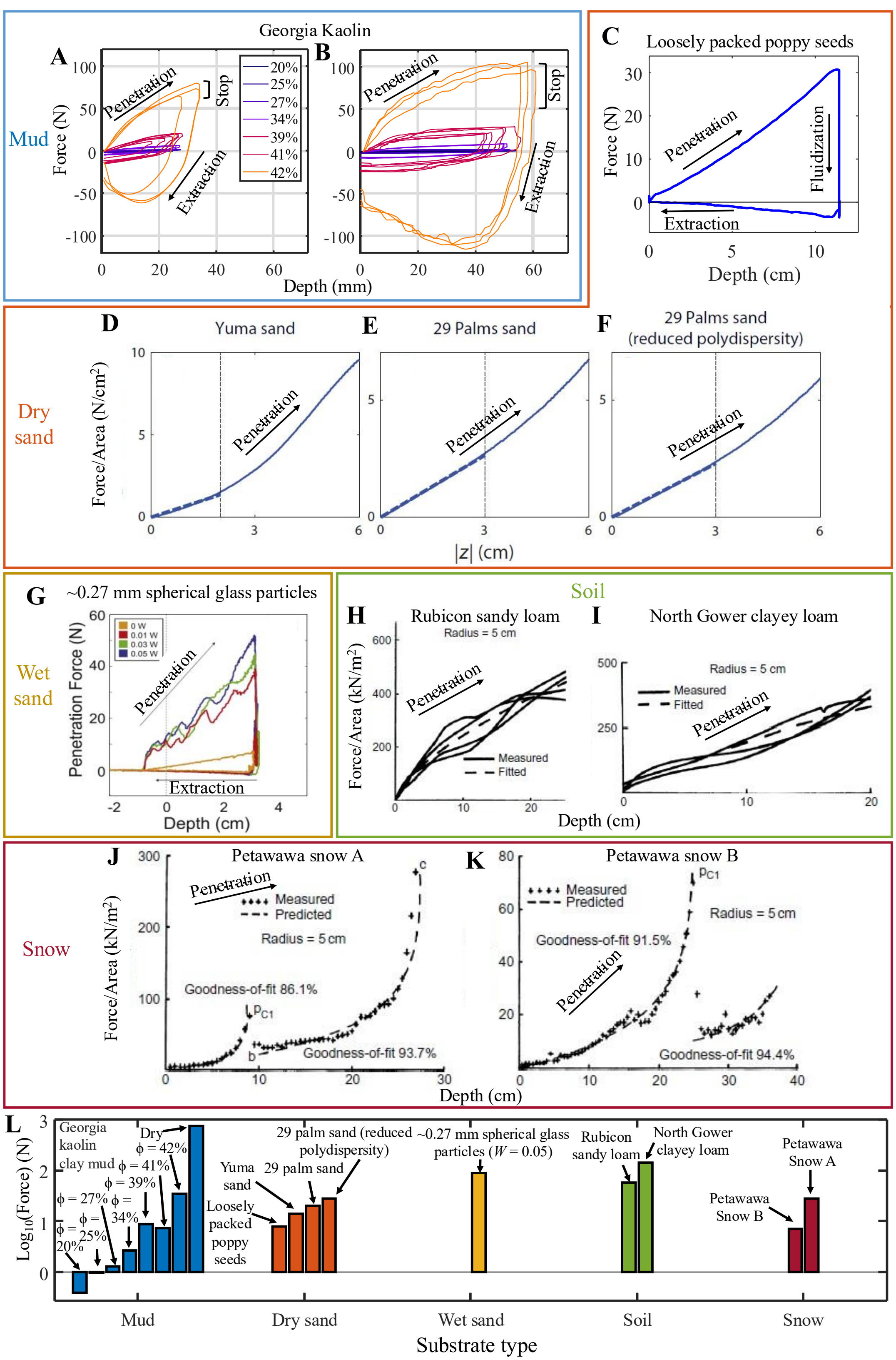

**Fig. 5. Comparison of vertical force during penetration and extraction and yield strength in resisting penetration between mud and other flowable substrates.** Vertical force as a function of depth: **(A, B)** Mud (Georgia Kaolin clay). **(C–F)** Dry sand of various kinds. **(G)** Wet sand. Water content $W = 0$ (loosely packed, $\phi = 0.58$), 0.01, 0.03, and 0.05 is defined as the ratio of mass of water to mass of solid particles. **(H–I)** Soil. **(J–K)** Snow. Both snow samples have an ice layer. C–F, G, and H–K are reproduced from (Li et al., 2013), (Sharpe et al., 2015), and (Wong, 2009), respectively. **(L)** Comparison of yield strength in resisting penetration across all substrates above, measured by vertical force calculated for a horizontal disc of 28.3 $cm^2$ area at a 0.64 cm depth during downward penetration, assuming force is proportional to area. Note the logarithmic scale to compare across three orders of magnitude.

In addition, the large downward force during extraction due to mud sticking to the intruder (Figs. 3A, 5A, B) means that mud can exert large downward (and backward) pulling force against any upward motion of an animal, robot, or vehicle (Artoni et al., 2019; Chen et al., 2024; Das et al., 1994; Hertog, 2017; Shin et al., 1994). Therefore, an animal, robot, or vehicle can become bogged down more easily in mud than in other flowable substrates.

A consequence of this is that locomotion on mud also induces substantially larger mechanical energy lost to the substrate during the extraction phase, when an animal or robot pulls its body and/or appendages upward and forward. Because force integrated over displacement is mechanical work or energy, the large downward (negative) force during extraction contributes to about the same mechanical energy lost to mud as compared to that on sand (with negligible downward force).

### 3.8. Penetrometers are useful for other flowable substrates

Mud is difficult to characterize using commercial penetrometers (Sec. 2.9), because it is not sensitive enough for weaker mud and cannot measure force as a function of depth that provides useful information (Sec. 3.7). Our custom-made penetrometers (Fig. 2D–E) are sensitive enough even for the weakest clay mud in the solid–fluid transition regime, have a large enough range for drier mud (whose strength is comparable to other flowable substrates, Sec. 3.7), and can measure force as a function of depth during both intrusion and extraction. These capabilities make them useful for other flowable substrates.

### 3.9. Penetration test to characterize mud yield strength is more nuanced

The nonlinear dependence of vertical force on depth makes it more nuanced to characterize mud yield strength than for dry sand using penetration tests. In dry sand, due to a hydrostatic-like pressure, vertical force is basically proportional to depth (Li et al., 2013). Thus, a horizontal disc penetrated to any depth gives basically the same force per unit depth, which, when further normalized to intruder area, measures the slope of yield stress vs. depth due to the hydrostatic-like pressure. Thus, commercial penetrometers that only go to a fixed depth are suitable. In mud, however, force increases with depth nonlinearly. Moreover, this nonlinear relationship is not the same for mud of different $\phi$ (Fig. 3A, i–vii). This highlights the inadequacy of commercial penetrometers that only penetrate to a fixed depth and the usefulness of our custom penetrometers capable of measuring force as a function of depth.

Interestingly, for our clay mud, measuring forces at different depths results gave basically the same dependences of mud strength on $\phi$ (Fig. S4), except that a higher depth gave a higher signal-to-noise ratio.

### 3.10. Characterization in the deep penetration limit is needed

The nuance of mud strength characterization may also result from another of its potential differences from dry sand. Recent constitutive modeling using frictional plasticity predicted that, unlike cohesionless flowable substrates like dry sand (Agarwal et al., 2021b; Li et al., 2009; Li et al., 2013) and weakly cohesive substrates such as wet sand (Sharpe et al., 2015), strongly cohesive media like clay mud (and muddy substrates with a high clay fraction $\eta$) should not have a hydrostatic-like pressure ((Agarwal et al., 2023), Supplementary Information, Section S2). If this is true, the dependence of vertical force on depth that we observed is likely from a “surface boundary effect” in the “shallow” penetration limit, where penetration depth is smaller than or comparable to intruder size (Peng et al., 2009). Future experiments should test this by measuring whether penetration force is independent of depth at depths far exceeding intruder size (i.e., in the "deep" penetration limit). Characterization spanning these two limits together will inform biological and engineering systems ranging from those moving on the surface to those burrowing deep into the substrates.

### 3.11. Characterizing non-vertical intrusion, stickiness, rate-dependence, and hysteresis is needed

Our characterization of yield strength during downward penetration is only a first step towards systematic characterization of mud properties and understanding how they challenge locomotion, for several reasons.

First, forces and flow during localized intrusion in flowable substrates are highly complex, and vertical penetration and extraction of a horizontal disc only captures one aspect of the challenges relevant to locomotion. We need methods suitable for mud of a wide range of wetness to measure other types of forces, such as during horizontal drag (Sotelo et al., 2022), rotational shear (Shakeel et al., 2020), and resistive forces on intruders of arbitrary shapes and trajectories (Agarwal et al., 2023; Kerimoglu et al., 2025; Li et al., 2013; Maladen et al., 2009; Treers et al., 2021), to comprehensively characterize the how muddy substrates resist localized intrusion.

In addition, yield strength, defined only for penetration but not extraction, does not capture how the stickiness of strongly cohesive muddy substrates affects forces and how this varies with wetness. Our visual observations of mud sticking to an intruder during extraction showed this (Sec. 3.1). In addition, our companion animal study (Ramesh et al., submitted) showed that mud stickiness strongly affects locomotion and that this was affected by changes in mud wetness in complex ways. Drier mud was not very sticky (e.g., $\phi$ = 39%, Georgia Kaolin clay mud), and there was a minimal amount of mud sticking to the animal and pulling it downward and backward as it extracted its body and fins. As mud became wetter ($\phi$ = 34%, Georgia Kaolin clay mud), it became weaker and stickier. As a result, the mudskipper sank more and during stance the contact area of the animal's body and fins with mud increased. More mud stuck to them and pulled them backward and downward as the animal extracted them. On the wettest and weakest mud tested ($\phi$ = 27%, Georgia Kaolin clay mud), mud stickiness level (as measured by how rapidly the mud sticking to the intruder disc detached during extraction) actually reduced, but because the animal sank much more, the amount of mud sticking to it still increased, resulting in the largest downward and backward pulling during extraction. Overall, the animal had difficulty in crutching forward not only from increased drag due to larger sinkage but also increased downward and backward pulling forces from mud sticking to it. This became so much of a problem that it even induced the animal to thrust its tail to assist crutching or even to jump in order to move (Ramesh et al., submitted). Therefore, we need better ways to quantify not only how easy mud sticks to an intruder, but also how this coupled with intruder sinkage changes the overall pulling forces, and how these vary with wetness.

Moreover, forces and flow in mud are rate-dependent (Coussot, 2017) (more so than in dry sand (Maladen et al., 2009)) and history-dependent (hysteresis) (Coussot, 2017) (likely stronger than in dry sand (Schiebel et al., 2020) due to stronger cohesion). This has been studied more for bulk mud flow (Coussot, 2017), but less for localized intrusion into mud relevant to locomotion. Here we only focused on intrusion at a fixed low speed (~0.3 cm/s) and a single cycle of intrusion and extraction, considering the slow vertical speed of the animal in our companion animal study (Ramesh et al., submitted). Future studies should more comprehensively characterize these effects during localized intrusion.

### 3.12. Summary

We classified granular solid–water mixtures into sandy and muddy substrates, each with sub–categories depending on how much water there is. Our comprehensive locomotion literature review in this context (Table 1) revealed the importance to control, vary, and characterize strongly cohesive mud and a lack of methods compared to those for cohesionless dry sand and weakly cohesive wet sand. We developed methods to control and vary wetness, and accurately characterize the yield strength of muddy substrates, which can also track its drift over a long time. Specifically, we developed an automated mud mixer to efficiently prepare uniform mud of different wetness (Sec. 2.3, Fig. 2A) and a sealing method to minimize water evaporation during storage (Sec. 2.7, Fig. 2C). Because commercial penetrometers are not sensitive enough for weaker mud or characterize its complex force–depth relationship informative of substantial locomotor challenges of mud, we developed two custom penetrometers for stationary and mobile use (Sec. 2.9–2.13, Fig. 2D–E). Our methods produced spatially uniform mud, with minimal temporal drift of yield strength (~8%) over ~1 hour typical of locomotion experiments (Sec. 3.3–3.4, Fig. 3C–E). Over many days, mud became gradually drier and substantially stronger (~2 times) from water evaporation over many weeks (Sec. 3.5, Fig. 4), but this is much slower than if mud is left open. Our force measurements (Sec. 3.7) showed that mud (except when dried out) is much weaker (1–2 orders of magnitude) and sticks much more than other flowable substrates to induce pulling force during extraction. Thus, animals and robots can much more easily get bogged down on mud.

**Acknowledgements**

We thank Luke Moon, Mia Urban, Kapi Ketan Mehta, and Jiangqi Tan for help with mud strength characterization experiments, Lucas An and Milla Ivanova for mud preparation assistance, Ken Kamrin, Feifei Qian, and Philippe Coussot for discussion, and two anonymous reviewers for suggestions.

**Competing Interests**

The authors declare no competing or financial interests.

**Author contributions**

Conceptualization: CL; methodology: DR, QF, GS, ZS, JR, CL; validation: DR, QF, GS, ZS; investigation: DR, GS, QF; data curation: DR; writing: DR, CL; visualization: DR, CL; supervision, project administration, funding acquisition: CL.

**Funding**

This study was funded by the Burroughs Wellcome Fund Career Award at the Scientific Interface, a Johns Hopkins University Bridge Grant, and an National Science Foundation Foundational Research in Robotics grant.

**Data availability**

Codes for real time control and data acquisition for the portable manual and stationary automatic penetrometer are available at https://doi.org/10.6084/m9.figshare.30984856.

**Supplementary figures**

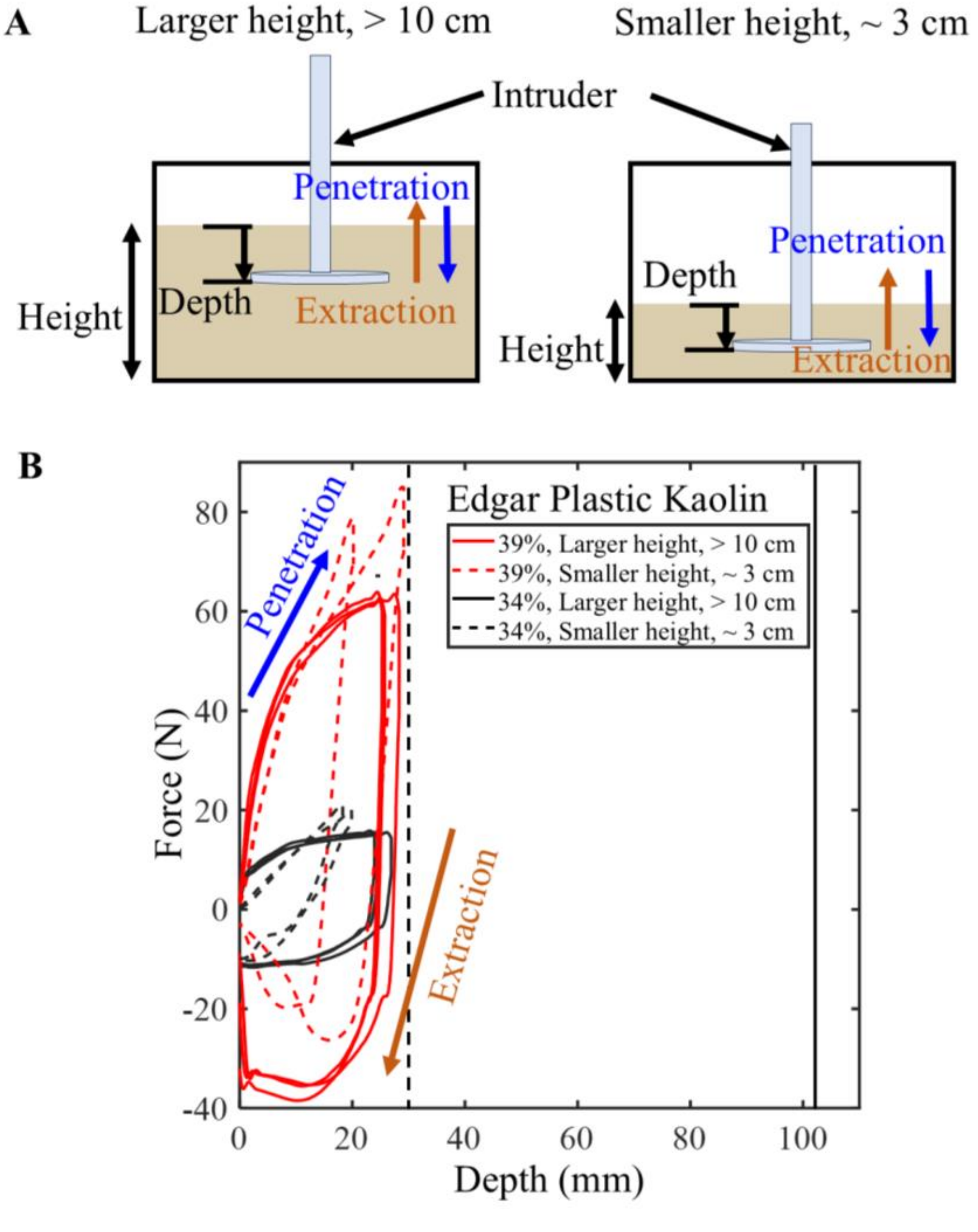

**Fig. S1. Boundary effects from the bottom of container. (A)** Schematics of a horizontal disc penetrated into mud filled to a large vs. a smaller height (10 vs. 3 cm, shown by the solid and dashed vertical lines in B). **(B)** Vertical force as a function of depth in mud with a larger (solid) vs. a smaller (dashed) height, for Edgar Plastic Kaolin clay mud of $\phi$ = 34% (black) and 39% (red). Each curve of the same treatment is a different trial. Note the different force vs. depth relationship between force in mud of a larger and that of a smaller height. In particular, force increases quickly with depth towards the end of penetration in mud of a smaller height (dashed vertical black line), which is absent in mud of a larger height (solid vertical black line).

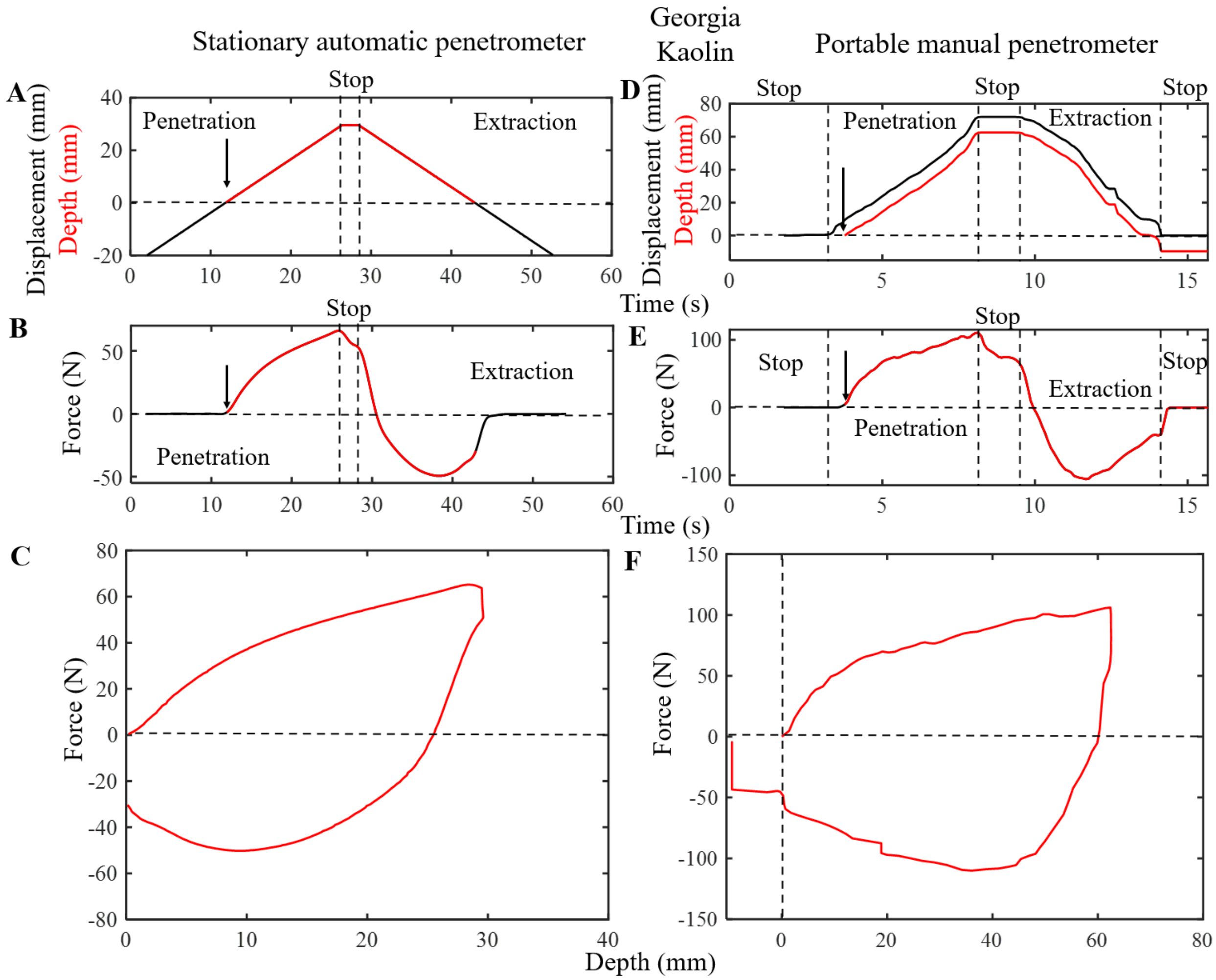

**Fig. S2. Data processing for penetration tests**. A–C and D–F are for stationary automatic and portable manual penetrometers, respectively, using Georgia Kaolin clay mud with $\phi$ = 42%. **(A, D)** Downward vertical displacement (black) and depth (red) as a function of time. Depth is obtained by offsetting displacement to zero at initial contact with the mud surface, which is determined when force is 0.03 N and 0.3 N for A and D, respectively. **(B, E)** Vertical force as a function of time. **(C, F)** Vertical force as a function of depth. Arrow in A, B, D, E show when force is zero on the mud surface.

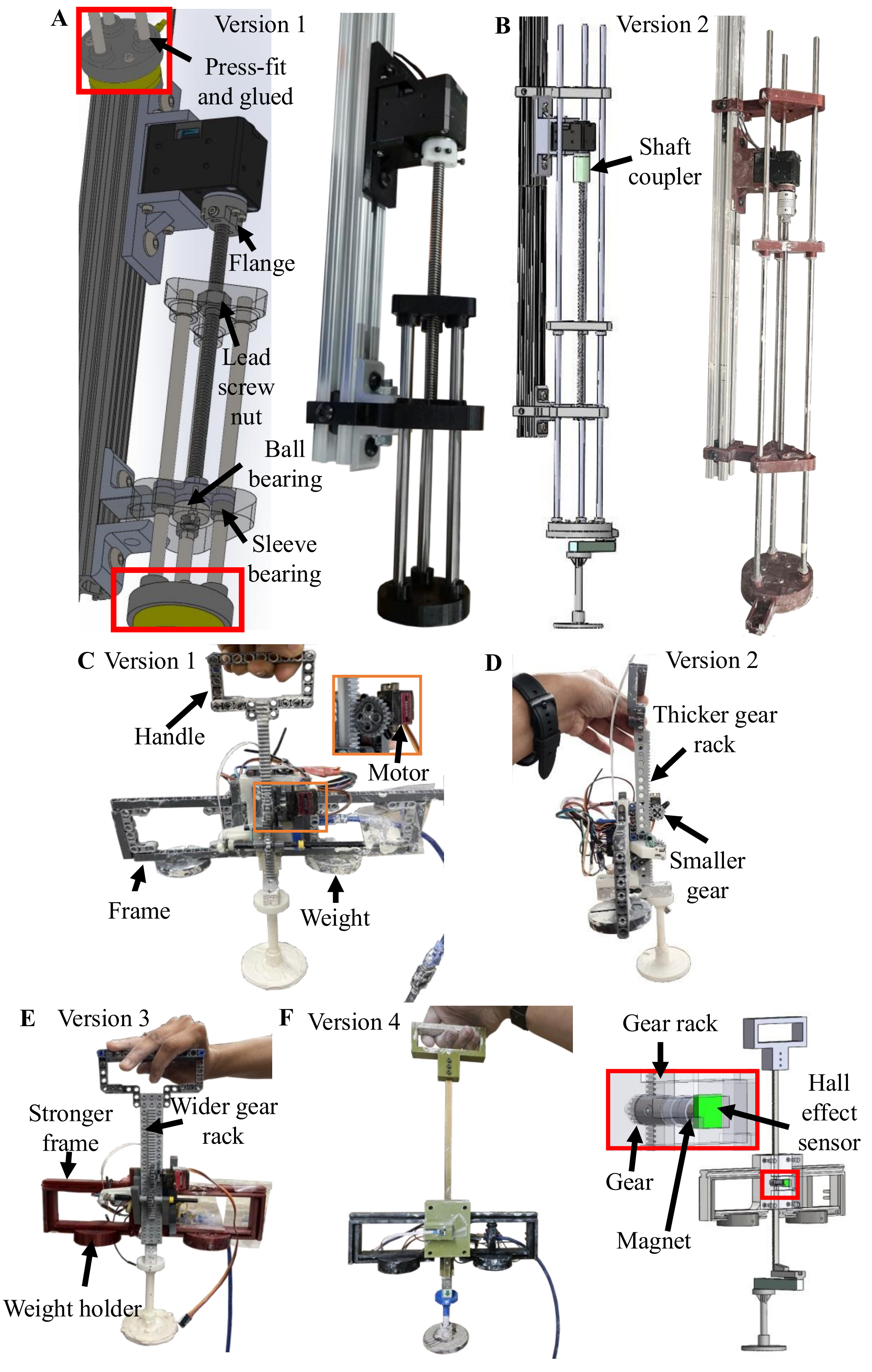
A
Version 1
Press-fit
and glued
Flange
Lead
screw
nut
Ball
bearing
Sleeve
bearing
B
Version 2
Shaft
coupler
C Version 1
Handle
Motor
Frame
Weight
D
Version 2
Thicker gear
rack
Smaller
gear
E Version 3
Stronger
frame
Wider gear
rack
Weight holder
F Version 4
Gear rack
Hall
effect
sensor
Gear
Magnet

**Fig. S3. Iterative improvements of custom penetrometers. (A, B) Stationary automatic penetrometer. (A)** Version 1 used a shorter threaded rod which reached only a shallower depth. **(B)** Version 2 (final version) used a longer threaded rod to increase the maximal reachable depth. It used two 3D-printed parts mounted onto the T-slotted framing from both ends to prevent linear slider from wobbling during use. It also used a shaft coupler due to difficulty in aligning the motor shaft with screws accurately using of a 3D-printed flange in Version 1. **(C–F) Portable manual penetrometer. (C)** Version 1 had a handle and frame for good support during use. **(D)** Version 2 had a thicker gear rack to prevent bending of gear rack due to the plastic material of the LEGO parts. It used a smaller gear which provides insufficient resistance when pushing into mud with lower $\phi$. **(E)** Version 3 had a wider gear rack support to prevent side-to-side movement. It also had a stronger 3-D printed frame with a better weight holder, because the frame made up of LEGO parts in Version 1 was attached with hot glue which wore off easily. **(F)** Version 4 (final version). It had a metal gear rack and a metal gear with a set screw, which secured a linear motion shaft (all four parts from McMaster-Carr, Princeton, NJ, USA) that was attached to a 3D-printed part holding the magnet. Weights added to the penetrometer helped keep the intruder disc perpendicular to mud surface.

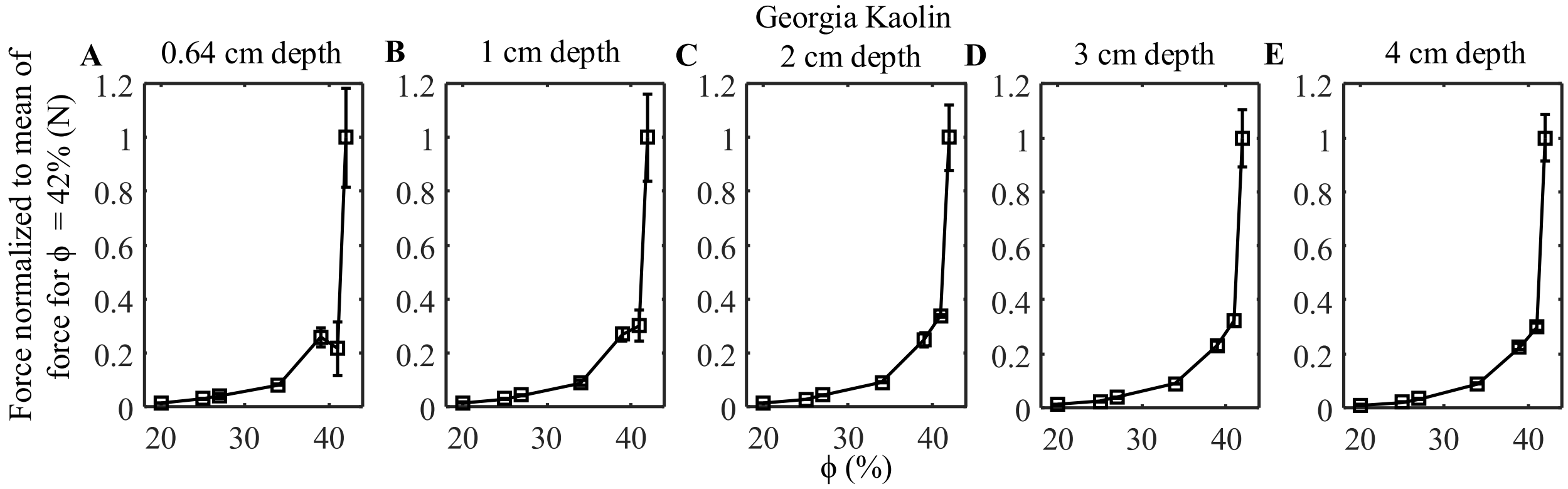

**Fig. S4. Vertical force during downward penetration vs. ϕ, comparing across different depths at which force is measured.** For each depth, average forces measured for each ϕ is normalized to the average force of that depth for ϕ = 42%. Error bars show ± s.d. of 3 trials for each treatment. Georgia Kaolin clay mud was used in these tests.